\documentclass[fleqn,usenatbib]{mnras}

\usepackage{newtxtext,newtxmath}

\usepackage[T1]{fontenc}
\usepackage{ae,aecompl}

\usepackage{graphicx}	
\usepackage{amsmath}	
\usepackage{amssymb}	
\usepackage{multicol} 
\usepackage{multirow}
\usepackage{comment}
\usepackage{array}
\usepackage{soul}
\hypersetup{colorlinks=true,citecolor=blue,linkcolor=purple,filecolor=black,runcolor=black,breaklinks=true}
\usepackage{etoolbox}
\makeatletter
\patchcmd\@combinedblfloats{\box\@outputbox}{\unvbox\@outputbox}{}{%
   \errmessage{\noexpand\@combinedblfloats could not be patched}%
}%
 \makeatother
\usepackage{ae,aecompl}
\usepackage[usenames]{color}
\usepackage{hyperref}
\usepackage{threeparttable}
\usepackage{mathrsfs}
\usepackage{scalerel}

\definecolor{purple}{RGB}{160,32,240}

\definecolor{red}{RGB}{225,50,50}

\newcommand{\HST}{\emph{HST}}
\newcommand{\JWST}{\emph{JWST}}
\newcommand{\Spitzer}{\emph{Spitzer}}
\newcommand{\Muv}{\ensuremath{\mathit{M}_{\mathrm{UV}}^{ }}}
\newcommand{\Muvth}{\ensuremath{\mathit{M}_{\mathrm{UV}}^{\mathrm{th}}}}
\newcommand{\Msol}{\ensuremath{\mathrm{M}_{\odot}}}

\newcommand{\muv}{\ensuremath{m_{\mathrm{UV}}^{ }}}
\newcommand{\muvth}{\ensuremath{m_{\mathrm{UV}}^{\mathrm{th}}}}
\newcommand{\muvthcen}{\ensuremath{m_{\mathrm{UV}}^{\mathrm{th,cen}}}}

\newcommand{\Mpch}{\ensuremath{\mathrm{Mpc}\ h^{-1}}}
\newcommand{\logten}{\ensuremath{\log_{10}}}
\newcommand{\Mst}{\ensuremath{\mathrm{M}^{\ast}}}

\newcommand{\Mhalo}{\ensuremath{\mathrm{M_{halo}}}}
\newcommand{\Mhalolog}{\ensuremath{\mathrm{log}\left(\mathrm{M_{halo}}/\mathrm{M}_{\odot}\right)}}

\newcommand{\Mpeak}{\ensuremath{\mathrm{M_{peak}}}}

\newcommand{\sigmahalo}{\ensuremath{\sigma_{\mathrm{halo}}}}

\newcommand{\zphot}{\ensuremath{\mathrm{z}_{\mathrm{phot}}}}

\newcommand{\thetamax}{\ensuremath{\theta_{\mathrm{max}}}}

\newcommand{\Lya}{\ensuremath{\mathrm{Ly}\alpha}}

\newcolumntype{P}[1]{>{\centering\arraybackslash}p{#1}}

\title[Clustering with JWST]{Clustering with JWST: Constraining Galaxy Host Halo Masses, Satellite Quenching Efficiencies, and Merger Rates at z=4-10}

\author[R. Endsley et al.]{
Ryan Endsley$^{1}$\thanks{E-mail: rendsley@email.arizona.edu},
Peter Behroozi$^{1}$,
Daniel P. Stark$^{1}$, 
Christina C. Williams$^{1,2}$,
\newauthor{
Brant E. Robertson$^{3,4}$,
Marcia Rieke$^{1}$,
Stefan Gottl\"ober$^{5}$,
Gustavo Yepes$^{6,7}$}
\\
$^{1}$Steward Observatory, University of Arizona, 933 N Cherry Ave, Tucson, AZ 85721 USA\\
$^{2}$NSF Fellow\\
$^{3}$Department of Astronomy and Astrophysics, University of California, Santa Cruz, 1156 High Street, Santa Cruz, CA 95064 USA\\
$^{4}$Institute for Advanced Study, 1 Einstein Drive, Princeton, NJ 08540 USA\\
$^{5}$Leibniz-Institut f\"{u}r Astrophysik, D-14482 Potsdam, Germany\\
$^{6}$Departamento de F\'{i}sica Te\'{o}rica, M\'{o}dulo 8, Facultad de Ciencias, Universidad Aut\'{o}noma de Madrid, 28049 Madrid, Spain\\
$^{7}$CIAFF, Facultad de Ciencias, Universidad Aut\'{o}noma de Madrid, 28049 Madrid, Spain
}

\date{Accepted XXX. Received YYY; in original form ZZZ}

\pubyear{2019}

\begin{document}
\label{firstpage}
\pagerange{\pageref{firstpage}--\pageref{lastpage}}
\maketitle

\begin{abstract}
Galaxy clustering measurements can be used to constrain many aspects of galaxy evolution, including galaxy host halo masses, satellite quenching efficiencies, and merger rates. We simulate \JWST{} galaxy clustering measurements at z$\sim$4$-$10 by utilizing mock galaxy samples produced by an empirical model, the \textsc{UniverseMachine}. We also adopt the survey footprints and typical depths of the planned joint NIRCam and NIRSpec Guaranteed Time Observation program planned for Cycle 1 to generate realistic \JWST{} survey realizations and to model high-redshift galaxy selection completeness. We find that galaxy clustering will be measured with $\gtrsim$5$\sigma$ significance at z$\sim$4$-$10. Halo mass precisions resulting from Cycle 1 angular clustering measurements will be $\sim$0.2 dex for faint (-18 $\gtrsim$ \Muv{} $\gtrsim$ -19) galaxies at z$\sim$4$-$10 as well as $\sim$0.3 dex for bright (\Muv{} $\sim$ -20) galaxies at z$\sim$4$-$7. Dedicated spectroscopic follow-up over $\sim$150 arcmin$^2$ would improve these precisions by $\sim$0.1 dex by removing chance projections and low-redshift contaminants. Future \JWST{} observations will therefore provide the first constraints on the stellar-halo mass relation in the epoch of reionization and substantially clarify how this relation evolves at z$>$4. We also find that $\sim$1000 individual satellites will be identifiable at z$\sim$4$-$8 with \JWST{}, enabling strong tests of satellite quenching evolution beyond currently available data (z$\lesssim$2). Finally, we find that \JWST{} observations can measure the evolution of galaxy major merger pair fractions at z$\sim$4$-$8 with $\sim$0.1$-$0.2 dex uncertainties. Such measurements would help determine the relative role of mergers to the build-up of stellar mass into the epoch of reionization.
\end{abstract}






\section{Introduction} \label{sec:introduction}

Over the past decade, multiple observational programs have opened a window into the first billion years following the Big Bang (see \citealt{Stark2016_ARAA} for a review). 
Ground and space-based photometric campaigns have revealed over one thousand galaxies at z$\gtrsim$6, enabling constraints on luminosity functions and star formation rate densities out to z$\sim$10 \citep[e.g.][]{Atek2015b,Bouwens2015_LF,Finkelstein2015_LF,Livermore2017,Ono2018,Oesch2018_z10LF} as well as galaxy stellar mass functions out to z$\sim$8 \citep[e.g.][]{Stark2009,Gonzalez2011,
Grazian2015,Song2016}. 
\JWST{} promises to advance luminosity and stellar mass function constraints at high redshifts via detailed spectra and extremely deep ($m \sim 29-30$) photometry in the near to mid-infrared.
However, less appreciated is \JWST{}'s potential to place new constraints on early galaxy evolution via high-redshift galaxy clustering measurements.

Galaxy clustering measurements are commonly used to infer halo masses by exploiting the strong relation between halo mass and clustering strength \citep[e.g.,][]{Mo1996,Tinker2010}. These halo masses can then be used to infer the stellar-halo mass relation, a fundamental constraint on the connection between galaxies and their host halos (see \citealt{Wechsler2018} for a review). Significant effort has been devoted to measuring z$\gtrsim$4 galaxy clustering, revealing that the most UV luminous galaxies at z$\sim$4$-$7 tend to reside in the most massive halos \citep{BaroneNugent2014,Harikane2016,Harikane2018,Ishikawa2017,Hatfield2018}. Combined with stellar mass constraints, high-redshift clustering results may already be signaling evolution in gas cooling efficiency, feedback efficiency, and merger rates from z$\sim$4 to z$\sim$7 \citep{Harikane2016,Harikane2018}. However, stringent ($\lesssim$0.5 dex error) halo mass constraints at z$\geq$4 are primarily limited to bright (\Muv{} $\sim$ -20) galaxies at z$\lesssim$5 \citep{BaroneNugent2014,Harikane2018}. It therefore remains unclear how the stellar-halo mass relation evolves at z$\gtrsim$6, leaving a gap in our understanding of galaxy evolution in the early Universe and, in particular, the epoch of reionization. 

\JWST{}'s greatly increased sensitivity is expected to significantly improve the precision of high-redshift clustering measurements. Deep (m$\gtrsim$29-30) NIRCam photometry from Cycle 1 programs will enable the detection of $\sim$5000 galaxies at z$>$6 and $\sim$300 at z$>$9 \citep{Mason2015,Williams2018}. Such increased numbers will highly benefit clustering measurements because Poisson noise falls as the linear inverse of galaxy number density \citep{Peebles1980,Landy1993}. Deeper imaging also enables the selection of lower mass halos which are less susceptible to cosmic variance \citep[e.g.,][]{Somerville2004,Trenti2008}. 

\JWST{}'s NIRSpec will also enable the first spatial galaxy clustering measurements at z$>$4 thanks to its multiplex capabilities and sensitivity to strong rest-optical lines at these redshifts (e.g,, H$\alpha$ and [OIII]; \citealt{Chevallard2019_NIRSpecSimulation,deBarros2019}). These spatial clustering measurements will avoid chance projections as well as low-redshift contaminants present in angular clustering measurements. 

All of these advancements should improve constraints on galaxy halo masses, and consequently, the stellar-halo mass relation at z$>$4. Recent studies have predicted the clustering of z$\sim$8-10 halos expected to host galaxies detectable with \JWST{} \citep{Bhowmick2018,Zhang2019_Clustering}. Adopting idealized assumptions, \citet{Zhang2019_Clustering} concluded that galaxy clustering should be measurable to $\sim$4$-$5$\sigma$ significance at z$\sim$10. Here, we simulate \JWST{} galaxy clustering measurements at z$\sim$4$-$10, noting that our methodology provides more observationally realistic predictions compared to \citet{Zhang2019_Clustering} in the following ways. First, we model high-redshift galaxy selection completeness by adopting the typical depths of a planned Cycle 1 program, the \JWST{} Advanced Deep Extragalactic Survey (JADES; \citealt{Williams2018}), and utilizing an empirical model, the \textsc{UniverseMachine} \citep{Behroozi2019}, to assign galaxy properties to halos from a dark matter simulation. Second, we simulate clustering measurements over the exact survey footprint of the JADES program, including detector and pointing gaps, to accurately account for boundary effects. Third, we account for peculiar motion distortions in spatial clustering measurements. Fourth and finally, we include satellite galaxies in our simulated measurements, which can significantly impact the recovered clustering signal because clustering strengths are density dependent. We furthermore simulate the process of inferring halo masses from these clustering measurements to provide the first predictions of z$\sim$4$-$10 halo mass precisions resulting from future \JWST{} surveys. 

We also investigate how \JWST{}'s improved sensitivity and spectroscopic capabilities will enable studies of satellite quenching and merger rates in the early Universe. Results from ground-based facilities have shown that satellites out to z$\sim$2 are systematically more quenched than field galaxies at fixed stellar mass \citep[e.g.,][]{vandenBosch2008,Kawinwanichakij2016}. These findings imply a strong environmental dependence on star-formation properties within the last $\sim$10 Gyr, possibly driven by the presence of a hot circumgalactic medium surrounding the host halo. We therefore test how well \JWST{} will facilitate the identification of high-redshift satellites and thereby push satellite quenching efficiency measurements to earlier epochs. We also investigate how \JWST{} will improve galaxy pair fraction measurements at z$>$4. Galaxy pair fractions can be used to infer galaxy merger rates and thus the relative contribution of mergers to stellar mass buildup throughout cosmic time \citep[e.g.,][]{Lotz2011,Mundy2017} as well as the AGN-merger connection \citep{Kocevski2012}. 

In \S\ref{sec:methods}, we describe our methods for generating realistic mock \JWST{} survey realizations (\S\ref{sec:methods_survey_realizations}), modeling high-redshift selection completeness (\S\ref{sec:methods_selection_effects}), calculating clustering strengths via the two-point correlation function (\S\ref{sec:methods_twoptCF}), and inferring halo masses from our simulated clustering measurements (\S\ref{sec:methods_halo_mass}). Our results are presented in \S\ref{sec:results_general} where we discuss the quality of our simulated z$\sim$4$-$10 galaxy clustering measurements (\S\ref{sec:results_clusteringStrengths}) and the halo mass uncertainties expected to result from these measurements (\S\ref{sec:results_MhaloPrecisions}), concluding with a discussion of how \JWST{} will improve constraints on the stellar-halo mass relation at z$\gtrsim$4. In \S\ref{sec:app_general}, we discuss how well \JWST{} observations will enable the identification of individual satellites (\S\ref{sec:app_satellites}) as well as measurements of galaxy major merger pair fractions (\S\ref{sec:app_merger}) at z$\gtrsim$4. Our main conclusions are listed in \S\ref{sec:conclusions}.

All magnitudes are quoted in the AB system \citep{OkeGunn1983}. Luminosity to stellar mass conversions assume a \citet{Chabrier2003} stellar initial mass function. We adopt a flat, $\Lambda$CDM cosmology with parameters ($\Omega_{\mathrm{M}} = 0.307$, $\Omega_{\mathrm{\Lambda}} = 0.693$, $\Omega_{\mathrm{B}} = 0.048$, $h=0.678$, $\sigma_8 = 0.823$, $n_s = 0.96$) consistent with \emph{Planck} results \citep{Planck2016}. Halo masses follow the \citet{Bryan1998} spherical overdensity definition and refer to peak historical halo masses extracted from the merger tree except where otherwise specified. All distances are quoted in comoving coordinates unless otherwise stated.

\section{Methods} \label{sec:methods}

In this section, we describe our methods for simulating z$\sim$4$-$10 \JWST{} galaxy clustering measurements and inferring halo masses from those measurements. We begin by detailing our procedure for generating realistic \JWST{} survey realizations based on a planned Cycle 1 GTO program, JADES \citep{Williams2018}, in \S\ref{sec:methods_survey_realizations}. We then model the selection of high-redshift galaxies to account for completeness, as well as low-redshift interlopers in the angular clustering measurements, as described in \S\ref{sec:methods_selection_effects}. Our methods for simulating galaxy clustering measurements are then detailed in \S\ref{sec:methods_twoptCF}, followed by a description of how we infer halo masses from these measurements in \S\ref{sec:methods_halo_mass}.

\subsection{Generating Mock JWST Survey Realizations} \label{sec:methods_survey_realizations}

Our approach to generating mock \JWST{} survey realizations begins by taking halo positions from a dark matter simulation so that predicted clustering strengths arise from $\Lambda$CDM theory. Here, we use the Very Small MultiDark Planck (VSMDPL\footnote{Details of VSMDPL can be found at \url{https://www.cosmosim.org/cms/simulations/vsmdpl/}. The series of MultiDark simulations is summarised at \url{https://www.cosmosim.org/cms/simulations/simulations-overview/}.}{}) dark matter simulation, which continues the series described in \citet{Klypin2016} to higher resolution. VSMDPL was run using \textsc{GADGET-2} \citep{Springel2005} with a box size of (160 \Mpch{})$^3$, containing 3840$^3$ particles with a high mass ($6.2\ \times\ 10^6\ \Msol{}\ h^{-1}$) and force (1 kpc $h^{-1}$) resolution. The cosmological parameters adopted in VSMDPL are from \citet{Planck2014} and listed in \S\ref{sec:introduction}. Halos and sub-halos within VSMDPL were identified using the \textsc{Rockstar} algorithm \citep{Behroozi2013_ROCKSTAR}, and merger trees were constructed using the \textsc{Consistent Trees} algorithm \citep{Behroozi2013_Trees}. 

We assign galaxy properties (including UV luminosities) to the VSMDPL halos using an empirical model, the \textsc{UniverseMachine} (hereafter referred to as the UM model; \citealt{Behroozi2019}). Briefly, the UM model utilizes observational constraints spanning z=0$-$10 to empirically infer halo star formation histories as a function of potential well depth, assembly history, and redshift. This procedure is performed using a Markov Chain Monte Carlo algorithm where each chain results in a mock catalog of galaxies paired with halos. In this work, we use the best-fitting catalog, hereafter referred to as the UM-VSMDPL mock catalog. Within the UM model, UV luminosities are computed from the star formation histories assigned to each halo using FSPS v3.0 \citep{Conroy2009_FSPS,Conroy2010_FSPS} and a \cite{Chabrier2003} stellar initial mass function, with dust attenuation scaled to match high-redshift observations \citep{Bouwens2016_ALMA}. 

\begin{figure}
\includegraphics{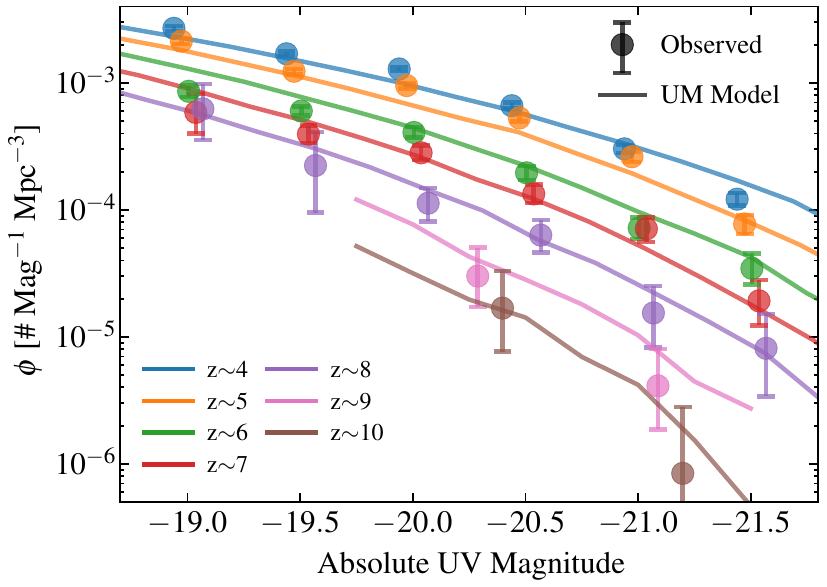}\par
\caption{Comparison of the z$\sim$4$-$10 UV luminosity functions from the entire mock galaxy catalog output by the \textsc{UniverseMachine} model (UM; \citealt{Behroozi2019}) used in this work (lines) versus the observational constraints input into that model. These observational data are taken from \citet{Finkelstein2015_LF} at z$\sim$4$-$8 and \citet{Bouwens2016_z910_LF} at z$\sim$9$-$10. Overall, the UV luminosity functions output by the model are well matched to input observations as expected.}
\label{fig:LF_comparison}
\end{figure}

At high redshifts, the galaxy-halo connection set by the UM model is primarily governed by the input empirical UV luminosity functions. These are taken from \citet{Finkelstein2015_LF} and \citet{Bouwens2016_z910_LF} at z$\sim$4$-$8 and z$\sim$9$-$10, respectively. As shown in Figure \ref{fig:LF_comparison}, the UV luminosity functions from the UM-VSMDPL mock catalog used in this work match the input constraints as expected. Additional high-redshift empirical constraints input into the UM model include UV-stellar mass relations (z=4$-$8), specific SFRs (z=0$-$8), and cosmic SFRs (z=0$-$9). See \citet{Behroozi2019} for additional details of the UM model including the calculation of UV luminosities. 

In Appendix \ref{appendix:abundance_matching}, we test how our results change if we instead adopt an abundance matching approach to assign UV luminosities to halos, finding no strong dependence. We also check whether our conclusions are dependent on the choice of empirical z$\sim$4$-$10 luminosity functions used to set the galaxy-halo connection at high redshifts. The only substantial difference occurs when adopting published z$\sim$10 luminosity functions that are $\sim$0.3 dex lower than those adopted by the UM model. This choice leads to significantly larger cosmic and Poisson variance in pair counts at this redshift, discussed further in Appendix \ref{appendix:UVLF_Uncertainty}.

To incorporate realistic sample variance into our simulated clustering measurements, we seek to only use mock galaxies that fall within the footprints of a planned Cycle 1 GTO program, the \JWST{} Advanced Deep Extragalactic Survey (JADES; \citealt{Williams2018}). Briefly, JADES will observe the GOODS-S and GOODS-N \HST{} Legacy fields with both photometry via NIRCam over 236 arcmin$^2$ and multi-object spectroscopy via NIRSpec over 142 arcmin$^2$. We therefore overlay the JADES footprints (including detector and pointing gaps) on mock survey volumes (i.e., lightcones) extracted from the UM-VSMDPL catalog. 

We choose to use only mock survey volumes that best match observed high-redshift galaxy number densities within the two GOODS fields to be observed by JADES. Our reasons for doing so are twofold. First, this procedure ensures our simulated clustering measurements will possess realistic Poisson variance which is of particular importance for the brightest and highest redshift samples. Secondly, it anchors the intrinsic clustering strengths of high-redshift galaxies within the GOODS fields because it has been shown that clustering strengths correlate with number density (i.e., environment; \citealt{Croton2007,Zehavi2018}). We generated 500 mock survey volumes in total by first extracting 100 lightcones from the UM-VSMDPL catalog, each spanning z=0$-$20 and subtending 50 arcminutes in right-ascension by 25 arcminutes in declination. Because each lightcone is large enough to encompass multiple JADES footprints for either field, we choose to extract five mock survey volumes from each lightcone centered at five evenly spaced RA positions within the lightcone. These mock volumes span an area of 0.4$\times$0.4 deg$^2$ to ensure that each can host a full footprint for either field.

We select the best matching mock survey volumes by first calculating the UV luminosity functions within each mock volume and then performing $\chi^2$ fits against observed luminosity functions within the two GOODS fields. Specifically, we adopt the luminosity functions reported by \citet{Finkelstein2015_LF} at z$\sim$4$-$8 and \citet{Oesch2014_z910LF} at z$\sim$10. To avoid fitting incomplete data, we only fit to the bright-end (\Muv{} $\leq$ -20) luminosity function data points which are $\gtrsim$50\% complete (see Figure 9 in \citealt{Finkelstein2015_LF}). We perform the fitting procedure for each GOODS field and each redshift interval individually. We also only use the central 0.2$\times$0.2 deg$^2$ of each mock volume as that better approximates the area covered by \HST{} in each GOODS field\footnote{Each mock survey volume has an area of 0.4$\times$0.4 deg$^2$ because parallel observations with JADES lead to more extended coverage relative to \HST{}.}{}.

One of the primary goals of this work is to determine 68\% confidence interval statistics for both clustering strength measurements and inferred halo masses. We find that generating 100 JADES realizations is sufficient for this purpose. Generating more realizations would require us to use mock survey volumes that are a poorer match to the empirical luminosity functions. To generate 100 realizations of the JADES program, we take the 10 best-fit mock survey volumes (out of 500) for each GOODS field and generate one realization for every possible pair of GOODS-N and GOODS-S volumes. The ten best-fit volumes for each field and redshift interval have reduced $\chi^2$ values less than three in all cases and are often $\lesssim$1.

\subsection{Modeling Selection of High-Redshift Galaxies} \label{sec:methods_selection_effects}

We now determine which mock galaxies would be used for high-redshift clustering measurements by simulating the selection of z$\sim$4$-$10 galaxies. In doing so, we adopt typical depths of the planned Cycle 1 program, JADES, and utilize empirically calibrated mock galaxy photometry and spectra produced by the JAdes extraGalactic Ultradeep Artificial Realizations (JAGUAR) package\footnote{We use the JAGUAR package because the UM model currently does not directly calibrate mock galaxy photometry and spectra.}{} \citep{Williams2018}. The overall goal here is to determine, as a function of redshift and UV magnitude, the photometric and spectroscopic selection completeness for angular and spatial clustering measurements, respectively.

For the angular clustering measurements, we assume that high-redshift galaxies will be photometrically selected using color cuts as commonly done with \HST{} imaging \citep[e.g.,][]{Stark2011,Gonzalez2012,Oesch2014_z910LF,Bouwens2015_LF,Bouwens2019_z910LF}. The color cuts adopted in this work are detailed in Appendix \ref{appendix:photz_selec} and are designed to separate galaxies into photometric redshift intervals of z=[3.7,4.4], [4.4,5.5], [5.5,6.7], [6.7,7.7], [7.7,9.0], and [9.0,11.0]. We will hereafter refer to these as the z$\sim$4, 5, 6, 7, 8, and 10 intervals, respectively. 

The photometric selection completeness, $\mathscr{C}_{P} (\mathrm{z},\Muv{})$, is calculated as the fraction of mock JAGUAR galaxies (as a function of redshift and UV magnitude) that are selected to lie within the redshift interval of interest. Prior to this mock selection process, we add noise to the photometry of each JAGUAR galaxy 50 times with noise set equal to the medium JADES/NIRCam depths \citep{Williams2018} for \JWST{} photometry and deep GOODS/ACS depths \citep{Bouwens2015_LF} for optical \HST{} photometry\footnote{Such \HST{} coverage is expected over a large fraction of the two GOODS fields to be imaged by JADES.}{}. We find that $>$50\% of z$\sim$4, 5, 6, 7, 8, and 10 galaxies with apparent rest-UV magnitudes of \muv{} = 28.0, 28.5, 28.8, 29.2, 29.2, and 29.5, respectively, are selected to lie within the correct photometric redshift interval. We further find that only using galaxies brighter than these magnitudes restricts the low-redshift (z$<$3) contamination fraction to reasonably small values ($\leq$15\%) within the context of JAGUAR. Therefore, mock galaxies in the UM-VSMDPL catalog fainter than the above magnitudes are ignored while brighter galaxies are selected at random for the simulated angular clustering measurements with probability equal to $\mathscr{C}_{P} (\mathrm{z},\Muv{})$.

Because NIRSpec micro-shutter assembly (MSA) targets must first be photometrically identified, the overall spectroscopic completeness, $\mathscr{C}_{S} (\mathrm{z},\Muv{})$, is determined by the spectroscopic redshift completeness, the photometric selection completeness, and the availability of MSA slits for targets. We assume that spectroscopic redshifts can be obtained from galaxies with at least one emission line detected at $\geq$5$\sigma$. Therefore, spectroscopic redshift completeness is calculated as the fraction of JAGUAR mock galaxies (as a function of redshift and UV luminosity) that have at least one emission line brighter than the 5$\sigma$ flux limit set by the planned medium JADES/NIRSpec depths. For the low-resolution (R$\sim$100) prism, the 5$\sigma$ limiting flux is 1.4$\times$10$^{-18}$ erg/s/cm$^2$ at 2.5$\mu$m. The medium-resolution (R$\sim$1000) G235M and G395M grisms are approximately twice as sensitive with 5$\sigma$ flux limits of 0.8 and 0.5$\times$10$^{-18}$ erg/s/cm$^2$ at 2.5$\mu$m and 4.5$\mu$m, respectively. Flux limits at other wavelengths are computed using the online sensitivity curves\footnote{https://jwst-docs.stsci.edu/display/JTI/NIRSpec+Sensitivity} for each disperser. We note that [OIII]$+$H$\beta$ equivalent widths in the JAGUAR catalog are consistent with inferences from observations \citep[e.g.][]{Labbe2013}. 

We assume that MSA targets will only include z$\gtrsim$4 photometric candidates that have an estimated $>$50\% spectroscopic redshift completeness and are brighter than the limiting photometric magnitudes summarized in Table \ref{tab:survey_params}. 
As detailed in Appendix \ref{appendix:spec_followup}, we find that the number of z$\gtrsim$4 photometric candidates will not exceed the number of available MSA slits assuming that the z$\sim$4-6 candidates will be targeted with the R$\sim$100 prism while the z$\gtrsim$7 candidates will be targeted with the R$\sim$1000 G235M and G395M grisms. Adopting those dispersers, we infer that z$\sim$4, 5, 6, 7, and 8 galaxies with \muv{} = 28.0, 28.5, 28.8, 29.0, and 28.7 are $>$50\% likely to have at least one emission line detection with NIRSpec. We therefore adopt these as the limiting magnitudes for simulating spatial clustering measurements where brighter galaxies are selected with probability equal to $\mathscr{C}_{S} (\mathrm{z},\Muv{})$. This net spectroscopic completeness is calculated as the convolution of the spectroscopic detection completeness and photometric selection completeness, both as a function of redshift and UV luminosity. This assumes that every z$\gtrsim$4 photometric candidate in a given NIRSpec pointing can eventually be placed on an MSA mask. In practice, it is likely that only $\sim$50\% of such sources can be placed on MSA masks in the Cycle 1 JADES program alone to avoid overlapping spectra. We discuss how a 50\% slit placement efficiency would impact our simulated spatial clustering measurements in Appendix \ref{appendix:spec_followup}.

We do not consider spectroscopic samples at z$\sim$10 because the very strong rest-optical lines such as [OIII] and H$\alpha$ are no longer accessible to NIRSpec at z$>$9, suggesting low spectroscopic completeness in this regime. While moderately strong lines such as [OII]$\lambda$3729 and [NeIII]$\lambda$3869 are still accessible to NIRSpec out to z$\sim$12, these lines are expected to be $\gtrsim$3$\times$ weaker than [OIII] at these redshifts \citep{Tang2018}. As such, we do not expect a sufficient number of z$\sim$10 galaxies to be spectroscopically detectable for spatial clustering measurements, at least within the medium JADES/NIRSpec survey. We discuss ways to address this challenge in \S\ref{sec:results_clusteringStrengths}. 

\begin{table}
\centering
\begin{tabular}{P{1cm}P{2cm}P{1.75cm}P{1.75cm}} 
\multicolumn{4}{c}{Adopted Limiting Rest-UV Apparent Magnitudes} \\
\hline
 & Targeted & Angular & Spatial\\
 & Redshift Interval & Clustering & Clustering\\
\hline
z$\sim$4 & 3.7$-$4.4 & 28.0 & 28.0\\
z$\sim$5 & 4.4$-$5.5 & 28.5 & 28.5\\
z$\sim$6 & 5.5$-$6.7 & 28.8 & 28.8\\
z$\sim$7 & 6.7$-$7.7 & 29.2 & 29.0\\
z$\sim$8 & 7.7$-$9.0 & 29.2 & 28.7\\
z$\sim$10 & 9.0$-$11.0 & 29.5 & -\\
\hline
\end{tabular}
\caption{The limiting rest-UV apparent magnitudes used to select high-redshift galaxies when simulating \JWST{} clustering measurements. All of the angular apparent magnitudes correspond to an approximate absolute magnitude of \Muv{} = -18. These magnitudes were determined by modeling high-redshift selection completeness using typical depths of the planned Cycle 1 JADES program as described in \S\ref{sec:methods_selection_effects}. We do not simulate spatial clustering measurements at z$\sim$10 because of the low spectroscopic completeness expected in this regime.} \label{tab:survey_params}
\end{table}

\subsection{Measuring Clustering Strengths Via the Two-Point Correlation Function} \label{sec:methods_twoptCF}

We simulate galaxy clustering measurements using the two-point correlation function as it allows us to consider the full scale-dependent clustering of galaxies. This approach is similar to recent studies \citep[e.g.,][]{Harikane2016,Harikane2018,Hatfield2018,Zhang2019_Clustering} and goes beyond past studies that have focused on a single-variable measurement of, e.g., the correlation length. The spatial two-point correlation function, $\xi(r)$, quantifies the excess pair counts at a given 3D separation distance, $r$, compared to pair counts for randomly-distributed galaxies \citep[e.g.,][]{Peebles1980}:
\begin{equation}
dP = n\left[1+\xi(r)\right] dV,
\end{equation}
where $dP$ is the number of excess pairs, $n$ is the galaxy number density, and $dV$ is the volume element. Because peculiar motions distort the observed line-of-sight distance, we instead measure the projected spatial correlation function,
\begin{equation}
\mathrm{w}_\mathrm{p} (r_\mathrm{p}) = \int_{-\pi_{\mathrm{max}}}^{\pi_{\mathrm{max}}} \xi(r_\mathrm{p}, \pi)\ d \pi.
\end{equation}
Here, the 3D separation distance, $r$, has been split into a projected 2D distance, $r_\mathrm{p}$, and a line-of-sight distance, $\pi$, that includes peculiar motion distortions. We adopt $\pi_{\mathrm{max}}$ values of 10, 7, 5, 5, and 3 \Mpch{} for the z$\sim$4, 5, 6, 7, and 8 bins, respectively. These values were chosen to encompass the majority of peculiar motion distortions, calculated via the halo velocity information in the UM-VSMDPL mock catalog. Our adopted $\pi_{\mathrm{max}}$ values also satisfy the resolution limitations of each NIRSpec disperser (see \S\ref{sec:methods_selection_effects}), assuming that emission line wavelength measurements can be made to within $\sim$1/4 of the resolution element.

We calculate w$_\mathrm{p}(r_\mathrm{p})$ using the \cite{Landy1993} estimator:
\begin{equation}
\mathrm{w}\left(D1, D2\right) = \frac{1}{R1R2} \left[D1D2 - 2\times D1R2 + R1R2\right],
\end{equation}
where $D1D2$ is the number of pairs between two galaxy samples ($D1$ and $D2$) at a given projected separation distance ($r_\mathrm{p}$) and $D1R2$ ($R1R2$) are the number of data-random (random-random) pairs, appropriately normalized (see \citealt{Landy1993}). To ensure that our random samples contribute negligible Poisson noise, we generate them using a surface density of $10^5$ objects per arcmin$^2$ which is more than 1000$\times$ that of our most populated mock galaxy sample. 

We simulate angular clustering measurements using the angular two-point correlation function, w($\theta)$. For this, we again use the \citet{Landy1993} estimator and a random sample with surface density of $10^5$ objects per arcmin$^2$. Because low-redshift galaxies contaminate high-redshift photometrically selected samples, we account for the resulting reduction in measured angular two-point correlation functions (at most 50\%) via the methods described in Appendix \ref{appendix:photz_contaminants}. 

We assume that galaxy clustering will be observationally measured in samples binned by redshift and threshold apparent rest-UV magnitudes, \muvth{} (e.g., \muvth{} = 29 means that only \muv{} $<$ 29 galaxies are included in the sample). Such binning is common in observational clustering studies at high redshifts \citep[e.g.,][]{Harikane2016, Harikane2018, Hatfield2018}. First, we adopt a threshold magnitude equal to the limiting magnitude at each redshift (see Table \ref{tab:survey_params}). These mock galaxy samples are auto-correlated ($D1 = D2$) and simply correspond to the entire sample used for our simulated clustering measurements at their respective redshift. We also consider galaxy samples with threshold magnitudes one or two magnitudes brighter than the limiting magnitude for redshifts where sufficient numbers of such bright galaxies are expected to lie within our adopted survey area. These brighter samples are cross-correlated with the entire galaxy sample at their respective redshift to minimize Poisson noise in pair counts. Table \ref{tab:sigmaMs} shows the full list of our adopted threshold magnitudes.

We simulate correlation function measurements for each \JWST{} survey realization described in \S\ref{sec:methods_survey_realizations}. Because Poisson error can be significant in any single mock clustering measurement, we also simulate correlation function measurements using all twenty mock survey volumes adopted for each redshift (see \S\ref{sec:methods_survey_realizations}). These `true' correlation functions are used to assess the accuracy of simulated JADES clustering measurements via the $\chi^2$ statistic,
\begin{equation}
    \chi^2 = \sum_i \frac{\left(\mathrm{w}_{i,\mathrm{meas}} - \mathrm{w}_{i,\mathrm{true}}\right)^2}{\sigma_i^2}.
\end{equation}
Here, w$_{i,\mathrm{meas}}$ is the simulated two-point correlation function measurement for a separation bin $i$, w$_{i,\mathrm{true}}$ is the `true' two-point correlation function at that separation, and $\sigma_i$ is the simulated jackknife error for w$_{i,\mathrm{meas}}$. We use jackknife errors because they are commonly adopted in observational clustering studies \citep[e.g.,][]{Harikane2016,Harikane2018,Coil2017} and we find that they will reasonably approximate a Gaussian error distribution (Appendix \ref{appendix:jackknife}). These errors are calculated from 10 jackknife samples of roughly equal area split by constant right-ascension over the JADES survey footprint. 

We also assess the significance of each simulated clustering measurement by comparing it to the null result (w=0). When calculating accuracy and significance, we exclude the innermost distance bin (i.e., the intrahalo term) which is dominated by satellite clustering. Satellite clustering, particularly for faint samples at high redshifts, remains highly uncertain in dark matter simulations due to the artificial loss of low-mass satellites \citep[e.g.,][]{vandenBosch2018a,vandenBosch2018b}. While the UM model compensates for these effects (see Appendix \ref{appendix:satellite_evolution}), we choose to exclude the intrahalo term to reduce any potential bias introduced by this correction method. 

We also report how the significances of our simulated clustering measurements change when instead utilizing a covariance matrix. As in \citet{Zhang2019_Clustering}, we calculate covariance matrices by computing correlation function measurements across 100 randomly selected mock survey volumes. While the covariance matrix provides a better representation of the true significance, it will likely not be possible to quantify this with \JWST{} observations as doing so requires observing a large number ($>$10) of independent fields. Here, we report both significances so that those obtained from a jackknife approach can be compared with those derived from a covariance matrix.

To avoid imposing artificial clustering signals, we ensure that the number densities of random samples, $n_{_{R}}$, used for spatial clustering measurements follow the redshift and luminosity evolution of galaxy number densities and their selection completeness:
\begin{equation}
n_{_{R}}(\mathrm{z}) \propto \int_{-\infty}^{\Muvth{}} \phi(\mathrm{z}, \Muv{})\ \mathscr{C}_S (\mathrm{z},\Muv{})\ d\Muv{}.
\end{equation}
Here, $\phi(\mathrm{z}, \Muv{})$ is the luminosity function taken from the best-fit redshift evolution reported by \citet{Finkelstein2015_LF}, $\mathscr{C}_S (\mathrm{z},\Muv{})$ is the spectroscopic completeness as a function of redshift and luminosity (see \S\ref{sec:methods_selection_effects}), and \Muvth{} is the corresponding value of \muvth{} at a given redshift, z.

\begin{figure} 
\includegraphics{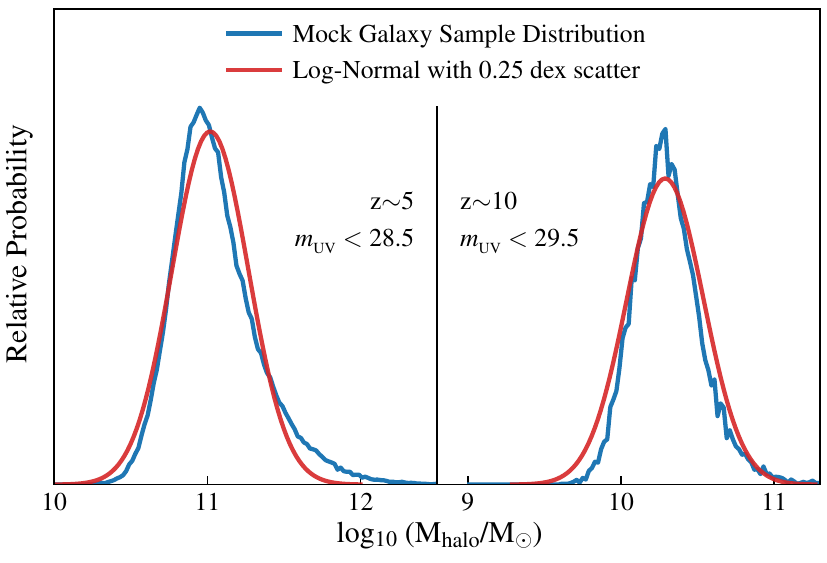}
\caption{Comparison of the true halo mass distribution of mock galaxy samples used to simulate \JWST{} clustering measurements (blue) and our adopted approximation to that distribution (red). As examples, we show this comparison for z$\sim$5 and 10 galaxy samples with \muv{} $<$ 28.5 and 29.5, respectively.}
\label{fig:CompareMhaloDistn}
\end{figure}

\subsection{Inferring Halo Masses} \label{sec:methods_halo_mass}

We now simulate the process of inferring halo masses from our simulated galaxy clustering measurements. By comparing these inferred halo masses to the true values in the UM-VSMDPL mock catalog, we are able to predict the precisions that will result from future \JWST{} clustering measurements. Specifically, we infer halo masses by fitting our simulated galaxy two-point correlation function measurements to a grid of correlation functions expected from halos at different masses. The inferred halo mass is then taken as the median value from the likelihood function, $\mathcal{L} \propto \mathrm{exp}(-\chi^2/2)$, where this process is performed for each mock galaxy sample and each survey realization separately. 

We first generate a grid of auto-correlation functions for halos of different masses.  Here, we use halo mass grid points between \Mhalolog{} = 9.50$-$12.75 with a spacing of 0.01 dex. To calculate correlation functions for each halo mass grid point, we auto-correlate samples of halos drawn from the twenty mock survey volumes used to simulate galaxy clustering measurements at the redshift of interest (see \S\ref{sec:methods_survey_realizations}). Halos are selected such that the typical (i.e., median) halo mass is equal to the grid point value of interest. They are also selected such that the resulting mass distribution is log-normal with 0.25 dex scatter, as this well-approximates the true halo mass distribution shape of all the z$\sim$4$-$10 mock galaxy samples considered in this work (see Figure \ref{fig:CompareMhaloDistn} for examples). To account for the partially random selection necessary to force this desired distribution, we perform the halo selection process ten times for each grid point and average the correlation functions from the ten selections. Halos are only selected if they lie within the redshift interval of interest (see Table \ref{tab:survey_params}). For each survey realization, we infer halo masses for the faintest galaxy sample in each redshift interval by fitting its simulated auto-correlation function with these gridded halo auto-correlation functions.

To infer halo masses for the brighter galaxy samples, we generate a grid of halo cross-correlation functions to match the procedure for simulating correlation function measurements of these brighter samples (\S\ref{sec:methods_twoptCF}). In this case, the halo samples $D1$ and $D2$ are selected to have different median masses $M_1$ and $M_2$, with the same 0.25 dex scatter as for the halo auto-correlation functions. $M_2$ is fixed to the inferred halo mass of the faintest galaxy sample for the redshift interval and survey realization of interest. The $M_1$ values (representing the brighter sample) are binned between $M_2$ minus 0.25 dex and \Mhalolog{} = 12.75, again with a spacing of 0.01 dex. As in the auto-correlation grid case, we average the correlation functions resulting from ten selections. We then fit for the value of $M_1$, which is taken as the inferred halo mass of the brighter sample.

Finally, we determine the true 68\% confidence interval halo mass precisions for each galaxy sample resulting from this simulated procedure. Specifically, the true halo mass precision, \sigmahalo{}, is defined such that the inferred halo mass is within \sigmahalo{} dex of the true typical halo mass for each mock galaxy sample (taken from the UM-VSMDPL catalog) in 68 of the 100 mock survey realizations. These halo mass precisions are therefore not necessarily equal to the halo mass uncertainties that would be inferred directly from observations. We find that observationally inferred halo mass uncertainties (68\% confidence intervals from the $\chi^2$ distribution using jackknife errors) can underestimate the true precision by up to a factor of $\sim$2 in dex. This is largely due to the systematic uncertainty imposed by the requisite assumption of the true halo mass distribution for the observed galaxy sample (see \S\ref{sec:results_MhaloPrecisions}) since we find that jackknife errors reasonably well capture the 68\% confidence interval from statistical fluctuations in the galaxy correlation function (Appendix \ref{appendix:jackknife}).

As motivated in \S\ref{sec:methods_twoptCF}, we use simulated jackknife errors when computing $\chi^2$ values because it will not be possible to derive accurate covariance matrices from observations alone. We do not calculate errors in our modeled halo mass-selected clustering strengths because the area from which these halos are selected is $>$10$\times$ larger than the mock observed fields thereby substantially reducing Poisson error. Furthermore, we exclude the intrahalo term when calculating $\chi^2$ values due to the model dependency of satellite clustering strengths (see \S\ref{sec:methods_twoptCF}). 

Finally, throughout this work, we use the peak halo mass (i.e., the maximum halo mass throughout that halo's lifetime prior to the redshift of interest) rather than the current halo mass\footnote{The peak halo mass is nearly equal to the current halo mass for isolated halos but may be significantly larger for satellites or close pairs where tidal stripping can lead to mass loss \citep{Reddick2013}.}{}. We do so because peak halo masses better predict halo clustering strengths \citep[e.g.,][]{Reddick2013,Guo2016}. As such, when calculating projected spatial halo correlation functions, we use random samples ($R1$ and $R2$) with number densities that follow the redshift evolution of the z=4$-$10 peak halo mass function from the UM-VSMDPL catalog (Appendix \ref{appendix:HMF}). The peak halo mass is on average $<$0.05 dex higher than the current halo mass for the z$\sim$4$-$10 mock galaxy samples used in this work.

\begin{figure*}
\begin{multicols}{2}
\includegraphics{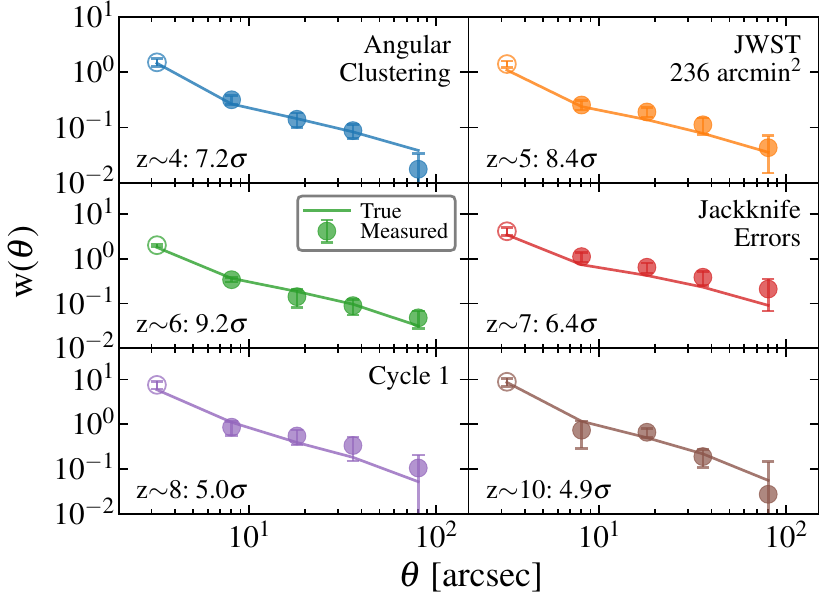}\par
\includegraphics{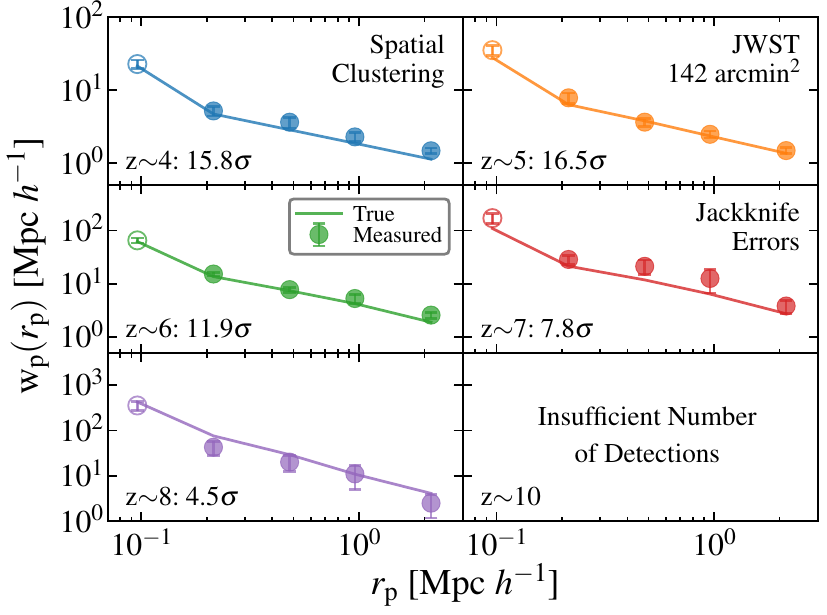}\par
\end{multicols}
\caption{Typical simulated angular (left) and projected spatial (right) galaxy two-point correlation function measurements at z$\sim$4-10 using the footprints and depths of the planned \JWST{} Cycle 1 GTO program JADES. Measurements (markers) are the result of auto-correlating the entire mock galaxy sample selected to lie within the redshift interval of interest. The typical significance (using jackknife errors) from 100 simulated survey realizations is listed in the lower left of each panel (one per redshift bin). The true correlation functions (see \S\ref{sec:methods_twoptCF}) are shown with a solid line to illustrate accuracy. We do not consider spatial clustering measurements at z$\sim$10 due to the low spectroscopic completeness expected in this regime (\S\ref{sec:methods_selection_effects}). The quality of spatial clustering measurements shown may require spectroscopic follow-up in addition to Cycle 1 observations due to possible target placement restrictions (see Appendix \ref{appendix:spec_followup}).}
\label{fig:AutoCorrelations}
\end{figure*}

\section{Results}\label{sec:results_general}

We now present z$\sim$4$-$10 galaxy clustering measurements and halo mass precisions expected to result from future \JWST{} observations. In \S\ref{sec:results_clusteringStrengths}, we discuss the quality of simulated z$\sim$4$-$10 galaxy clustering measurements which adopt the footprints and typical depths of a planned Cycle 1 GTO program, JADES. Halo mass precisions resulting from these simulated clustering measurements are presented in \S\ref{sec:results_MhaloPrecisions}, followed with a discussion of how \JWST{} will improve constraints on the evolution of the stellar-halo mass relation at z$>$4. 

\subsection{z$\sim$4-10 Galaxy Clustering Measurements with JWST} \label{sec:results_clusteringStrengths}

We begin by investigating the quality of z$\sim$4$-$10 galaxy clustering measurements expected from future \JWST{} surveys. Figure \ref{fig:AutoCorrelations} shows simulated angular (left) and projected spatial (right) two-point correlation function measurements utilizing the footprint and depths of the planned Cycle 1 GTO program JADES \citep{Williams2018}. The median measurement significance from 100 survey realizations is shown in the lower left of each panel, one for each redshift bin. We find that Cycle 1 observations will enable $\gtrsim$5 sigma clustering measurements at z$\sim$4$-$10. 

Figure 3 shows examples of typical measurements expected with the JADES program. Specifically, we plot measurements from realizations with significances within 0.5 of the median significance and $\chi^2$ values (relative to the true clustering strengths; see \S\ref{sec:methods_twoptCF}) within unity of the median $\chi^2$. The median $\chi^2$ lies between 2.4$-$3.9 for the angular measurements and 2.7$-$7.7 for the spatial measurements, resulting in median reduced $\chi^2$ values that are less than two in all cases.

The high quality of z$\sim$4$-$10 angular clustering measurements expected from Cycle 1 observations are the result of NIRCam's $\sim$50$\times$ improved sensitivity relative to WFC3/\HST{}. Such capabilities will enable surveys that cover $\gtrsim$100 arcmin$^2$ regions with depths similar to \textit{Hubble} ultra-deep fields (m$\sim$29.5; \citealt{Bouwens2015_LF}). Access to galaxies $\sim$2 magnitudes further down the luminosity function will significantly reduce Poisson noise due to the steep faint-end slope \citep[e.g.,][]{Bouwens2015_LF,Finkelstein2015_LF}, and weaken cosmic variance because fainter galaxies reside in lower mass halos \citep[e.g.,][]{Somerville2004,Trenti2008,Moster2011}.

As shown in Figure \ref{fig:AutoCorrelations}, we find that the precision of Cycle 1 angular clustering measurements will gradually evolve with redshift. Specifically, the typical measurement significance will increase from $\sim$7$\sigma$ at z$\sim$4 to $\sim$9$\sigma$ at z$\sim$6, then decrease at higher redshifts to $\sim$5$\sigma$ at z$\sim$10. This evolution is caused by two sources of noise that have opposite trends with redshift. In general, Poisson noise increases with redshift because galaxy number densities approximately halve per unit redshift, at least out to z$\sim$8 \citep[e.g.,][]{Bouwens2015_LF,Finkelstein2015_LF}. Conversely, the noise caused by chance projections decreases with redshift because galaxies are more spare and more strongly clustered at higher redshifts. Noise from chance projections dominates at z$\lesssim$6 while Poisson noise begins to take over at higher redshifts.

\JWST{}'s NIRSpec will greatly reduce the impact of chance projections by efficiently delivering precise redshifts out to z$\sim$9 via the detection of strong rest-optical lines such as H$\alpha$ and [OIII] \citep[e.g.,][]{Chevallard2019_NIRSpecSimulation,deBarros2019}. This is of particular importance for the crowded z$\sim$4$-$5 samples where chance projections over hundreds of Mpc dilute strong clustering signals on sub-Mpc scales. Figure \ref{fig:AutoCorrelations} illustrates that NIRSpec is capable of delivering $\sim$15$\sigma$ spatial clustering measurements at z$\sim$4$-$5 over the 142 arcmin$^2$ JADES spectroscopic area, double the significance expected from angular measurements over the 236 arcmin$^2$ JADES NIRCam area. The spatial clustering measurements and significances shown in Figure \ref{fig:AutoCorrelations} assume spectroscopic follow-up of every z$\gtrsim$4 candidate brighter than the limiting magnitudes listed in Table \ref{tab:survey_params}. As discussed in Appendix \ref{appendix:spec_followup}, it is likely that only $\sim$50\% of such candidates can be targeted in Cycle 1 to avoid overlapping spectra. However, we choose to show results assuming 100\% follow-up completeness to illustrate results possible with dedicated spectroscopy in future cycles. These results are also possible with 50\% follow-up completeness given $\approx$4$\times$ the coverage\footnote{Poisson noise is proportional to 1/$\sqrt{D1D2}$ \citep{Landy1993}. Doubling galaxy number densities quadruples $D1D2$ and therefore reduces Poisson noise by a factor of 2. However, doubling the coverage with fixed number density only reduces Poisson noise by 1/$\sqrt{2}$.}.

We expect that angular clustering measurements will be more precise than spatial at z$\gtrsim$8 for the following three reasons. First, chance projections occur less often at higher redshifts because number densities drop and clustering strengths increase. Secondly, angular clustering measurements will suffer less Poisson and cosmic variance because of wider coverage. Finally, spectroscopic completeness declines at z$\gtrsim$8 as NIRSpec becomes less sensitive to strong [OIII] emission. We do not consider spatial clustering measurements at z$\sim$10 because NIRSpec's sensitivity to strong lines stops at z$\sim$9, suggesting low spectroscopic completeness in this regime as discussed in \S\ref{sec:methods_selection_effects}. It is possible that dedicated deep spectroscopic follow-up of z$\sim$10 photometric candidates would enable spatial clustering measurements at this redshift.

Notably, we find that angular clustering measurements will reach $\sim$5$\sigma$ significance at z$\sim$10 within Cycle 1, rivaling current clustering measurements at z$\sim$7 \citep{BaroneNugent2014}. This conclusion persists even when adopting the most pessimistic z$\sim$10 luminosity functions yet published so long as we take into account other planned Cycle 1 surveys. As detailed in Appendix \ref{appendix:UVLF_Uncertainty}, adopting the $\sim$0.3 dex lower z$\sim$9$-$10 luminosity functions reported by \citet{Bouwens2019_z910LF} and \citet{Oesch2018_z10LF} (relative to \citealt{Bouwens2016_z910_LF}, used for the UM model) decreases the typical z$\sim$10 angular clustering measurement significance from 4.9$\sigma$ to 3.0$\sigma$. However, this ignores the $\sim$0.8 mag increased depths within the 46 arcmin$^2$ deep JADES region \citep{Williams2018}. Cycle 1 observations will also include the 100 arcmin$^2$ Cosmic Evolution Early Release Science (CEERS; P.I. S. Finkelstein) program with NIRCam depths $\sim$0.4 mags shallower than those adopted in this work. Assuming z$\sim$10 Schechter parameters of $M_{\mathrm{UV}}^{\ast} = -20.60$ and $\alpha = -2.3$ \citep{Oesch2018_z10LF}, we estimate that combining all of these Cycle 1 observations will decrease Poisson noise by $\sim$60\% relative to the medium JADES program alone. Additional coverage provided by the CEERS program will also reduce cosmic variance. We therefore estimate that z$\sim$10 angular galaxy clustering will be measured at $\sim$5$\sigma$ significance with Cycle 1 surveys, even assuming low galaxy number densities at this redshift. 

When calculating the significances quoted above, we have omitted intrahalo clustering strength measurements due to model uncertainties in satellite clustering strengths (see Appendix \ref{appendix:satellite_evolution}). Including the intrahalo term generally boosts the predicted significances by $\sim$2$-$3$\sigma$. On the other hand, adopting a covariance matrix to remove correlations between different separation distances lowers our reported jackknife significances down to $\sim$6$-$7$\sigma$ at z$\sim$4$-$6 and $\sim$3$-$5$\sigma$ at z$\sim$7-10. This, however, does not impact our conclusions below on halo mass inferences because we adopted jackknife errors for this procedure.

\begin{figure*} 
\includegraphics{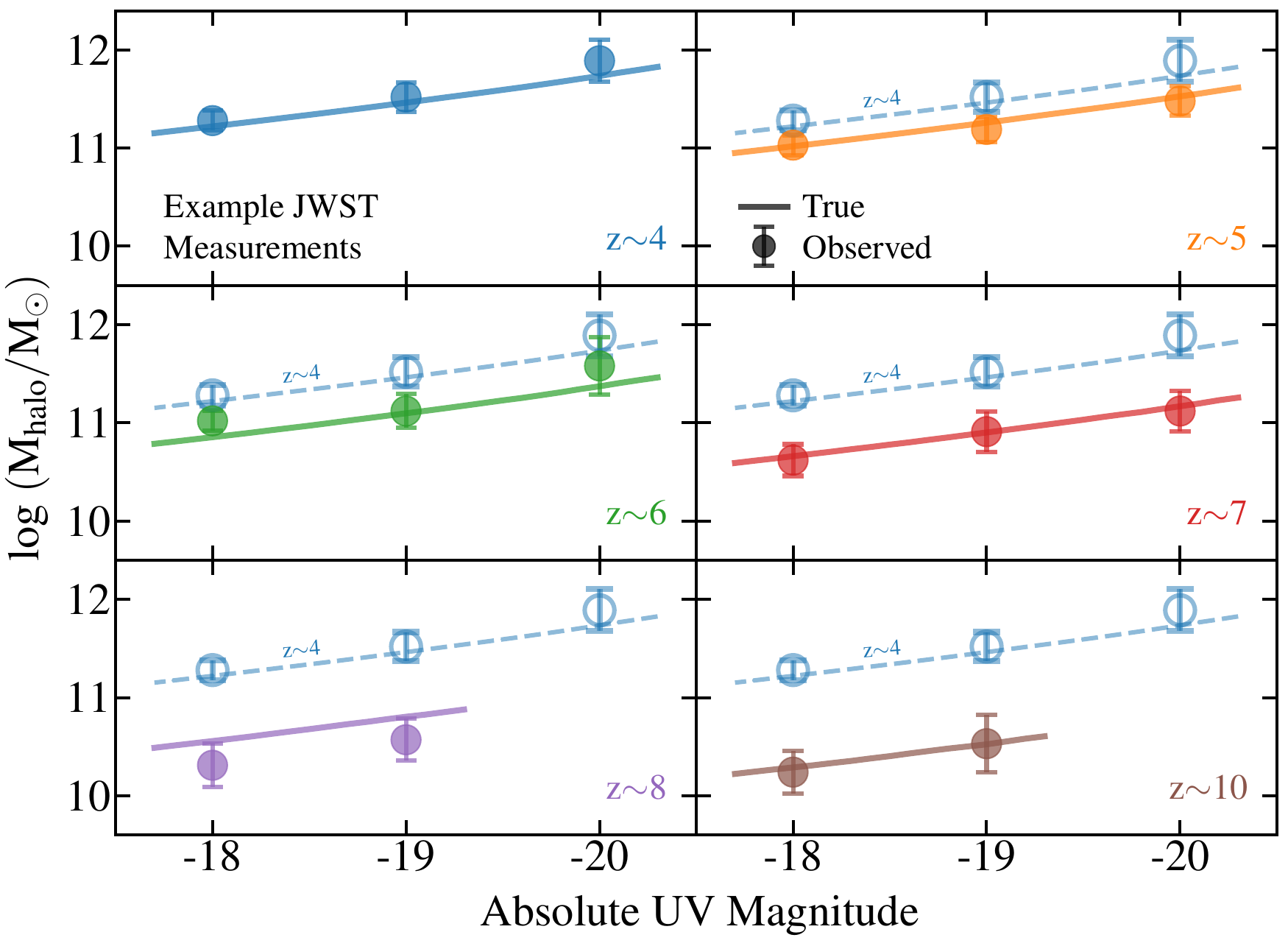}
\caption{An example simulated measurement of the \Muv{}$-$\Mhalo{} relationship inferred from Cycle 1 \JWST{} observations. Halo mass errorbars are taken from Table \ref{tab:sigmaMs} using spatial clustering precisions at z$\sim$4-6 and angular clustering precisions at z$\sim$7-10. On the x-axis, we use the integer absolute UV magnitude most closely corresponding to the threshold apparent magnitude of each galaxy sample. The true \Muv{}$-$\Mhalo{} relation from the UM-VSMDPL mock catalog is shown with a solid line at each redshift. The relation at z$\sim$4 is shown in each panel to illustrate the evolution in the \Muv{}$-$\Mhalo{} relation seen from z$\sim$4$-$10 in the simulated measurements.}
\label{fig:Mhalo_Muv_Measurement}
\end{figure*}

\subsection{Inferred Halo Mass Precision with \JWST{}} \label{sec:results_MhaloPrecisions}

Here, we predict the halo mass precisions that will result from \JWST{} galaxy clustering measurements at z$\sim$4$-$10 using the procedure outlined in \S\ref{sec:methods_halo_mass}. To briefly review, we define halo mass precisions, \sigmahalo{}, such that the inferred halo mass is within \sigmahalo{} dex of the true typical halo mass of the mock galaxy sample of interest in 68 of the 100 mock survey realizations. These values therefore represent the true errors in recovering halo masses, as opposed to the error estimates that would be obtained by marginalizing over observationally measured correlation function uncertainties obtained using, e.g., a jackknife approach.

In general, we find that z$\sim$4$-$10 halo mass precisions will be $\lesssim$0.25 dex. The implications of this precision are illustrated in Figure \ref{fig:Mhalo_Muv_Measurement} where we compare the inferred \Muv{}$-$\Mhalo{} relations (markers) with the true relation (line) taken from the entire UM-VSMDPL mock catalog\footnote{The volume of the UM-VSMDPL mock catalog is $\sim$500$\times$ that to be observed with the JADES program at each redshift.}{}. Halo masses within the UM-VSMDPL catalog decline by $\sim$1 dex from z$\sim$4 to z$\sim$10 at fixed UV luminosity. If such a strong evolution does indeed exist in the real Universe, our results suggest that it will be easily possible to measure this evolution from future \JWST{} observations.

As noted above, our reported precisions include both statistical and systematic components. The statistical component results from Poisson noise as well as field-to-field density fluctuations which alter the intrinsic clustering strengths at fixed halo mass. Angular clustering measurements also possess increased statistical uncertainties due to redshift contamination and chance projections. Combined, these statistical components contribute $\sim$0.05$-$0.15 dex to the halo mass uncertainties. Systematic uncertainty arises from the required assumption of the true halo mass distribution of the observed galaxy sample. An incorrect assumption of the distribution may bias the observationally inferred halo masses. To estimate the resulting systematic halo mass uncertainties, we compare the true correlation functions (\S\ref{sec:methods_twoptCF}) of each mock galaxy sample with the correlation functions of halos with a mass distribution following our adopted prescription (log-normal with 0.25 dex scatter) and median mass equal to that of the mock galaxy sample. These two correlation functions tend to differ by $\lesssim$0.1 dex for both our simulated angular and spatial clustering measurements, translating to $\sim$0.05-0.15 dex systematic uncertainties in the inferred halo masses of our galaxy samples. 

Table \ref{tab:sigmaMs} lists our predicted true halo mass precisions for each galaxy sample at z$\sim$4$-$10. In summary, we find that Cycle 1 angular clustering measurements will yield $\sim$0.2 dex halo mass precisions for faint (-18 $\gtrsim$ \Muv{} $\gtrsim$ -19) galaxies at z$\sim$4$-$10 as well as $\sim$0.3 dex precisions for bright (\Muv{} $\sim$ -20) galaxies at z$\sim$4$-$7. Dedicated spectroscopic follow-up of z$\sim$4$-$6 photometric candidates over the JADES/NIRSpec footprint would improve halo mass precisions by $\sim$0.1 dex at these redshifts. Specifically, spatial clustering measurements can yield high halo mass precision ($\sim$0.10$-$0.15 dex) for faint galaxies as well as moderate precision ($\sim$0.2 dex) for bright galaxies at z$\sim$4$-$6. 

\begin{table}
\centering
\begin{tabular}{ccccc}
\multicolumn{5}{c}{Predicted True Halo Mass Precisions with JWST}\\
\hline
 & \multicolumn{2}{c}{Angular Clustering} & \multicolumn{2}{c}{Spatial Clustering}\\
\hline
Redshift & \muvth{} & $\sigma_{\mathrm{halo}}$ & \muvth{} & $\sigma_{\mathrm{halo}}$ \\ 
\hline
\multirow{3}{*}{z$\sim$4} & 28.0 & 0.21 & 28.0 & 0.10 \\
                          & 27.0 & 0.21 & 27.0 & 0.15 \\
                          & 26.0 & 0.31 & 26.0 & 0.21 \\
\hline
\multirow{3}{*}{z$\sim$5} & 28.5 & 0.23 & 28.5 & 0.10 \\
                          & 27.5 & 0.15 & 27.5 & 0.13 \\
                          & 26.5 & 0.14 & 26.5 & 0.15 \\
\hline
\multirow{3}{*}{z$\sim$6} & 28.8 & 0.12 & 28.8 & 0.10 \\
                          & 27.8 & 0.17 & 27.8 & 0.17 \\
                          & 26.8 & 0.38 & 26.8 & 0.29 \\
\hline
\multirow{3}{*}{z$\sim$7} & 29.2 & 0.16 & 29.0 & 0.24 \\
                          & 28.2 & 0.21 & 28.0 & 0.30 \\
                          & 27.2 & 0.21 & 27.0 & 0.39 \\
\hline
\multirow{2}{*}{z$\sim$8} & 29.2 & 0.22 & 28.7 & 0.36 \\
                          & 28.2 & 0.21 & 27.7 & 0.38 \\
\hline
\multirow{2}{*}{z$\sim$10} & 29.5 & 0.22 & - & - \\
                          & 28.5 & 0.29 & - & - \\
\hline
\end{tabular}
\caption {Estimates of the true halo mass precisions, \sigmahalo{}, (in dex) expected to result from future \JWST{} clustering measurements at z$\sim$4$-$10. Each row shows the predicted precision for a given galaxy sample selected by redshift and threshold rest-UV magnitude, \muvth{}. These precisions are defined to reflect true 68\% confidence intervals (see \S\ref{sec:methods_halo_mass}) and result from adopting the footprints and typical depths of the \JWST{} Cycle 1 GTO program JADES. We do not consider spatial clustering measurements at z$\sim$10 due to the low spectroscopic completeness expected in this regime (\S\ref{sec:methods_selection_effects}).} \label{tab:sigmaMs}
\end{table}

Precisely inferring halo masses at high redshifts is of particular interest for settling the debate surrounding the evolution of the stellar-halo mass relation. This relation constrains the relative importance of various baryonic effects impacting galaxy evolution throughout cosmic time. Quantifying the evolution of this relation at high redshifts not only provides vital insight on the processes governing the early stages of galaxy evolution, but also the processes driving the ionizing output from galaxies during reionization. Many recent models have used observational data to infer the z$>$4 evolution of the stellar-to-halo mass relation with differing conclusions. Significant \citep{Behroozi2013_AbunMatch,Behroozi2015_HighZEvol,
Finkelstein2015_BaryonEfficiency,Harikane2016,Harikane2018,Sun2016}, moderate \citep{Moster2018}, and little evidence of evolution \citep{RodriguezPuebla2017,Stefanon2017_SHMR} have all been claimed. This discrepancy is largely due to currently poor empirical constraints on both halo and stellar masses at high redshifts.

Current halo mass uncertainties from clustering measurements (and independent of galaxy number density measurements) are $\gtrsim$0.5 dex at z$\sim$6-7 \citep{BaroneNugent2014} while no halo mass constraints yet exist at z$\geq$8. Our results therefore imply that future \JWST{} observations will provide markedly improved halo mass constraints at z$\sim$6-7 as well as the first constraints at z$\sim$8-10. We also expect z$>$4 stellar mass measurements to improve with \JWST{}. Such measurements currently suffer from poor sensitivity (m$\sim$25-26; \citealt{Bouwens2015_LF}), significant confusion \citep[e.g.,][]{Gonzalez2011,Song2016,Stefanon2017_SHMR}, and systematic uncertainties due to nebular contamination \citep[e.g.,][]{Stark2013_NebEmission,Song2016} in mid-infrared \Spitzer{}/IRAC photometry. NIRCam's $\sim$100-fold increase in mid-infrared sensitivity and greatly improved point-source resolution ($\sim$0.1 arcsec vs. $\sim$2 arcsec) will eliminate the first two challenges while NIRSpec spectroscopy and NIRCam medium band photometry will mitigate the last. We thus conclude that future \JWST{} observations will provide the first picture of the stellar-halo mass relation in the reionization era and substantially clarify how this relation evolves with redshift at z$>$4.

\section{Additional Applications With JWST} \label{sec:app_general}

High-redshift clustering measurements will enable many other studies relevant to understanding early galaxy evolution. In \S\ref{sec:app_satellites}, we discuss how well \JWST{} will enable inferences on satellite fractions and the identification of satellites and satellite-host pairs at z$\gtrsim$4. Such studies can be applied to inform models of sub-halo occupation and the environmental dependence on star formation in the early universe. In \S\ref{sec:app_merger}, we simulate Cycle 1 measurements of z$\gtrsim$4 galaxy pair fractions to assess how well \JWST{} will provide insight on galaxy merger rates and the relative role of mergers to stellar mass build-up throughout cosmic history.

\subsection{Constraining Galaxy Satellite Fractions and Quenching Efficiencies at z$\gtrsim$4} \label{sec:app_satellites}

Small-scale ($\lesssim$5 arcsec) galaxy clustering measurements can be used to infer galaxy satellite fractions. Inferences from current clustering measurements suggest that galaxy satellite fractions at z$\sim$4$-$6 increase from $\lesssim$1\% for very bright (\Muv{} $\lesssim$ -21) galaxies to $\sim$5\% for bright (\Muv{} $\sim$ -20) galaxies \citep{Ishikawa2017,Harikane2018,Hatfield2018}. They also suggest that z$\sim$4$-$6 galaxy satellite fractions decrease with redshift at fixed UV luminosity \citep[e.g.,][]{Harikane2018}. As illustrated in Figure \ref{fig:AutoCorrelations}, our results suggest that \JWST{} will deliver high-precision ($\sim$5$-$10$\sigma$) small-scale galaxy clustering measurements for faint (-18 $\gtrsim$ \Muv{} $\gtrsim$ -19) galaxies at z$\sim$4$-$10. The satellite fractions of these samples are assumed to be 10$-$15\% from the UM model (including orphan satellites; Appendix \ref{appendix:satellite_evolution}) with lower values at higher redshifts. We therefore predict that upcoming \JWST{} surveys will soon make it possible to verify that satellite fractions decrease at earlier times and increase in fainter samples at z$\sim$4$-$10. 

In addition to quantifying the overall fraction of satellites, \JWST{}'s spectroscopic capabilities will make it possible to determine, with high confidence, which z$\gtrsim$4 galaxies are satellites and which are hosts. This capability is of particular interest for testing the influence of environment on star formation in the early Universe. Studies at z$\lesssim$2 have found that satellites are systematically more quenched than field galaxies at fixed stellar mass where the quenching efficiency decreases with redshift \citep[e.g.,][]{Kawinwanichakij2016}. Empirically constraining satellite quenching efficiencies at z$\gtrsim$4 would test model predictions of when environment begins to strongly influence galaxy evolution and how rapidly this dependence set in.

To assist future studies in these endeavors, we provide optimal parameters for selecting z$\gtrsim$4 satellites and their hosts from future \JWST{} surveys. We assume that satellites and centrals will be observationally selected using an isolation criteria commonly adopted in lower-redshift studies \citep[e.g.,][]{Tal2013,Kawinwanichakij2014}. Specifically, we assume that a galaxy will be observationally identified as a central if it is the brightest galaxy (in the rest-UV) out to a specified maximum angular distance, \thetamax{}, and inferred line-of-sight distance $\pi_{\mathrm{max}}$. We further assume that galaxies will be identified as satellites if they lie within \thetamax{} and $\pi_{\mathrm{max}}$ of an inferred central galaxy brighter than some limiting magnitude, \muvthcen{} where that central is inferred to be the satellite's host. 

We simulate this selection process using mock galaxies from the UM-VSMDPL catalog (\S\ref{sec:methods_survey_realizations}) and compare the resulting satellite and host galaxy populations to the true populations within that catalog. Because the most significant impact of the host circumgalactic medium occurs within the virial radius \citep{Zhang2019_IGM}, UM-VSMDPL halos are defined as true satellites if they reside within the virial radius of a more massive halo which is defined as the true host. We account for spectroscopic completeness as described in \S\ref{sec:methods_selection_effects} only considering mock galaxies brighter than the limiting spectroscopic magnitudes listed in Table \ref{tab:survey_params}. We fix $\pi_{\mathrm{max}}$ values to 10, 7, 5, 5, and 3 \Mpch{} at z$\sim$4, 5, 6, 7, and 8, respectively (\S\ref{sec:methods_twoptCF}), corresponding to a redshift uncertainty of $\sigma_z \approx 0.02$. Because it will likely be difficult to observationally distinguish faint satellites from their hosts at separations of $\lesssim$2$\times$ the host half-light radius, we conservatively ignore all potential satellites within 0.5\arcsec of a brighter neighbor. This threshold value was chosen assuming a typical half-light radii of 0.15\arcsec for very bright (\Muv{} = -21) galaxies at z$\sim$4 \citep{Shibuya2015}. 

For each redshift interval, we test various maximum angular separations, \thetamax{}, and central galaxy limiting magnitudes, \muvthcen{}, seeking to optimize the number of correctly identified satellite and satellite-host pairs while keeping the fraction of contaminants to $<$20\%. We find that the parameter values listed in Table \ref{tab:BestSatIdenParams} are optimal. We choose a threshold 20\% contamination fraction because it is close to the lowest value that our mock procedure suggests will be possible with \JWST{} (particularly for satellite-host identification) while still providing large samples of true satellites and satellite-host pairs. Statistical background subtraction procedures utilized in lower redshift studies \citep[e.g.,][]{Tal2012,Kawinwanichakij2014} can be used to correct for contaminants.

Using the parameters listed in Table \ref{tab:BestSatIdenParams}, we find that, on average, $\approx$0.65, 1.05, 0.45, 0.15, and 0.05 true satellites will be identifiable per arcmin$^2$ at z$\sim$4, 5, 6, 7, and 8, respectively. We also find that $\approx$0.30, 0.75, 0.35, 0.15, and 0.05 true satellite-host pairs will be identifiable per arcmin$^2$ at z$\sim$4, 5, 6, 7, and 8, respectively, again on average. The number of identifiable true satellite-host pairs is lower than the number of identifiable true satellites because a galaxy can be the satellite of another, more massive halo that may not necessarily be the brightest due to scatter in the observed \Muv{}$-$\Mhalo{} relation.

These numbers suggest that Cycle 1 observations from the JADES and CEERS programs ($\sim$200 arcmin$^2$ of photometric and spectroscopic coverage) will enable the identification of $\sim$200 satellites\footnote{Here, we are assuming a NIRSpec MSA target placement efficiency of 50\% for Cycle 1 observations (Appendix \ref{appendix:spec_followup}).}{} at z$\sim$4$-$5. Studies of satellite quenching efficiencies at z$\sim$2 have used $\sim$450 satellites \citep{Kawinwanichakij2014,Kawinwanichakij2016}, suggesting that Cycle 1 observations will begin pushing constraints on the environmental impact of star formation\footnote{We assume that it will be possible to separate star-forming and quenched galaxies using spectral energy distribution fits of \JWST{} photometry and spectra to estimate rest-frame U-V and V-J colors as done in lower-redshift studies \citep{Kawinwanichakij2016}.} to z$\sim$5. Furthermore, given that current extragalactic \HST{} legacy surveys cover $\sim$750 arcmin$^2$ \citep{Grogin2011}, our results suggest that comprehensive follow-up of these regions with \JWST{} would provide z$\sim$6-8 satellite quenching constraints comparable to what is currently available at z$\sim$2. We conclude that \JWST{} is capable of testing how satellite quenching efficiencies continue to evolve with both redshift and host star formation efficiency at z$\gtrsim$4 and into the epoch of reionization.

It is worth discussing model dependencies of these conclusions. The UM model introduces orphan satellites (galaxies which no longer have an identifiable host halo in a dark matter simulation; \citealt{Wang2006}) to correct for the artificial disruption of low-mass satellites in simulations (Appendix \ref{appendix:satellite_evolution}). Because orphan satellites constitute approximately 50\% of satellites at z$\sim$4$-$8 within the UM-VSMDPL catalog, the accuracy of this satellite correction method impacts the numbers quoted above. In the worst-case scenario that no orphans should be introduced, we would be overestimating the number of identifiable z$\gtrsim$4 satellites and satellite-host pairs by a factor of $\approx$2. Fortunately, independent empirical constraints of z$\sim$4$-$5 galaxy major merger pair fractions suggest that the high-redshift orphan populations introduced by the UM model are reasonable (Appendix \ref{appendix:satellite_evolution}). It is also possible that z$\sim$4$-$8 galaxy number densities are underestimated in the UM-VSMDPL catalog due to the empirical luminosity functions adopted by the UM model (Appendix \ref{appendix:UVLF_Uncertainty}). If this is the case, we would expect the number of identifiable z$\gtrsim$4 satellites and satellite-host pairs quoted above to rise.

\begin{table}
\centering
\begin{tabular}{P{0.7cm}P{0.7cm}P{0.7cm}P{1cm}|P{0.7cm}P{0.7cm}P{0.85cm}}
\hline
 & \multicolumn{3}{c}{Satellites} & \multicolumn{3}{c}{Satellite-Host Pairs}\\
\hline
Redshift & \thetamax{} (\arcsec) & \muvthcen{} & \# per arcmin$^2$ & \thetamax{} (\arcsec) & \muvthcen{} & \# per arcmin$^2$ \\ 
\hline
z$\sim$4 & 4.25 & 27.5 & 0.65 & 3.50 & 26.0 & 0.30\\
\hline
z$\sim$5 & 3.75 & 28.0 & 1.05 & 3.25 & 27.0 & 0.75\\
\hline
z$\sim$6 & 3.25 & 28.3 & 0.45 & 3.00 & 27.3 & 0.35\\
\hline
z$\sim$7 & 2.75 & 28.5 & 0.15 & 2.25 & 28.0 & 0.15\\
\hline
z$\sim$8 & 2.50 & 28.2 & 0.05 & 2.50 & 27.7 & 0.05\\
\hline
\end{tabular}
\caption {Optimal parameters (\thetamax{} and \muvthcen{}; defined in \S\ref{sec:app_satellites}) to use when identifying satellites and satellite-host pairs (both the satellite and its specific host galaxy) at z$\sim$4$-$8 with $>$80\% confidence from spectroscopic \JWST{} surveys. We also show the expected surface number of true identifiable satellites and satellite-host pairs using these parameters. Here, we have modeled spectroscopic completeness using the methods described in \S\ref{sec:methods_selection_effects}.} \label{tab:BestSatIdenParams}
\end{table} 

\begin{figure} 
\includegraphics{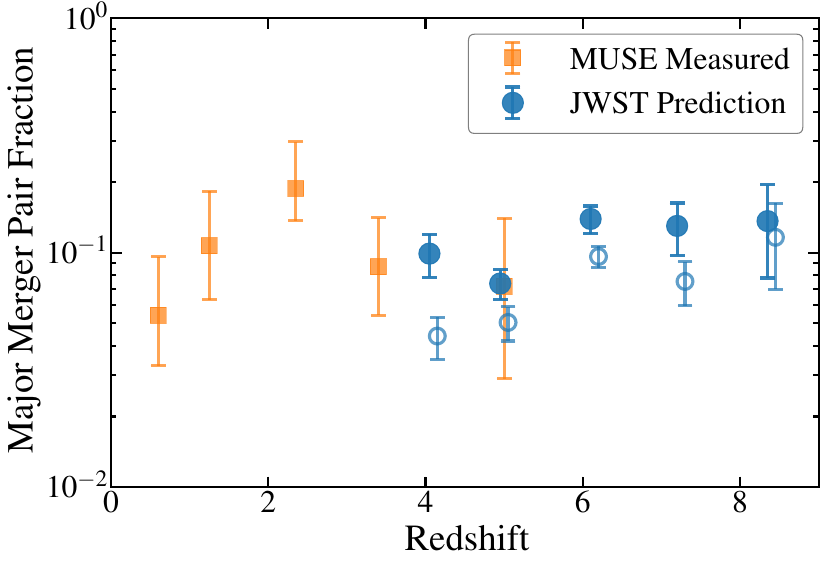}
\caption{Simulated \JWST{} measurements of the major merger galaxy pair fractions at z$\sim$4$-$8 (blue) using the footprints and typical depths of the planned Cycle 1 GTO program, JADES. We also plot for comparison the MUSE measurements from \citet{Ventou2017} from z$\sim$0$-$6 (orange). We find that future \JWST{} surveys will substantially improve current galaxy pair fraction measurements at z$\sim$4$-$6 and establish the first measurements at z$\sim$7$-$8. We also show simulated pair fraction measurements ignoring all orphan satellites (Appendix \ref{appendix:satellite_evolution}) with empty markers. We find that including orphans leads to better agreement with current observations.}
\label{fig:MajorMergerPairFraction}
\end{figure}

\subsection{Measuring Major Merger Galaxy Pair Fractions at z$\gtrsim$4} \label{sec:app_merger}

We now assess how well close galaxy pair fractions at z$\gtrsim$4 can be measured from future \JWST{} surveys. Pair fraction measurements can be applied to infer galaxy merger rates \citep[e.g.,][]{Patton1997,Kartaltepe2007} and the relative contribution of mergers to the build-up of stellar mass throughout cosmic history \citep[e.g.,][]{Mundy2017}. Identifying close pairs also enables tests on the connection between mergers and AGN activity \citep[e.g.,][]{Kocevski2012}.

The Multi Unit Spectroscopic Explorer (MUSE; \citealt{Bacon2010}) recently enabled the first spectroscopic galaxy pair fraction measurements at z$\sim$4-6 \citep{Ventou2017}. Because of \JWST{}'s much wider field of view and ability to detect strong rest-optical lines to z$\simeq$9, we expect that upcoming \JWST{} surveys will deliver improved galaxy pair fractions measurements at z$\gtrsim$4. We test this hypothesis directly by simulating z$\gtrsim$4 galaxy pair fraction measurements using the same mock volumes and survey footprints described in \S\ref{sec:methods_survey_realizations} which reflect the planned Cycle 1 program, JADES \citep{Williams2018}. We also account for spectroscopic completeness using the methods outlined in \S\ref{sec:methods_selection_effects}. 

\defcitealias{Ventou2017}{V17}

We assume that galaxy pair fractions will be measured with \JWST{} in a similar manner as described in \citet{Ventou2017}, hereafter \citetalias{Ventou2017}, which follows the methodology of \citet{deRavel2009} and \citet{LopezSanjuan2013}. Specifically, we identify a close pair as a system of two mock galaxies within a projected distance range of $r_\mathrm{p}^{\mathrm{min}} \leq r_\mathrm{p} \leq r_\mathrm{p}^{\mathrm{max}}$ and a rest-frame relative velocity of $\Delta \mathrm{v} \leq \Delta \mathrm{v_{max}}$. As in \citetalias{Ventou2017}, we adopt $r_\mathrm{p}^{\mathrm{max}}$ = 25 h$^{-1}$ kpc. The $r_\mathrm{p}^{\mathrm{min}}$ value is set equal to the projected distance corresponding to 0.5\arcsec at the redshift of interest to account for the fact that it will be difficult to distinguish faint galaxies within $\sim$2 half-light radii of a very bright neighbor (c.f. the 0.15\arcsec\ half-light radii of extremely luminous z$\sim$4 galaxies; \citealt{Shibuya2015}). Rather than use the $\Delta \mathrm{v_{max}}$ = 500 km s$^{-1}$ value from \citetalias{Ventou2017}, we adopt $\Delta \mathrm{v_{max}}$ = 1000 km s$^{-1}$ guided by the expected resolving power of JADES NIRSpec observations\footnote{As noted in \S\ref{sec:methods_selection_effects}, we assume R$\sim$100 at z$\sim$4-6 and R$\sim$1000 at z$\sim$7-8. The MUSE resolving power is R$\sim$3000 for \Lya{} at z$\sim$4$-$6.}. We do not expect that this increased choice of $\Delta \mathrm{v_{max}}$ will significantly impact the fraction of identified close pairs that will eventually merge (see, e.g., Figure 5 of \citealt{Patton2000}). For comparison, $\Delta \mathrm{v_{max}}$ = 1000 km s$^{-1}$ corresponds to $\pi_{\mathrm{max}}\approx$ 8 and 6 \Mpch{} at z=4 and 8, respectively. 

We further consider only major mergers, which we define as galaxy pairs with a stellar mass ratio $\leq$1/4 to keep in convention with many other pair fraction studies \citep[e.g.,][]{Man2016,Snyder2017,Mantha2018,Duncan2019}. Observed stellar masses are calculated by perturbing the true stellar mass of each mock galaxy (computed from the full star formation histories of the halo from the UM model) with a fixed scatter reflecting observational uncertainties. We assume that stellar mass uncertainties with \JWST{} will be 0.2 dex at z$\sim$4$-$8 though our conclusions are not significantly altered if we instead adopt an 0.3 dex uncertainty. 

Similar to \citetalias{Ventou2017}, we compute the major merger pair fraction, $f_{\mathrm{MM}}$, as
\begin{equation}
    f_{\mathrm{MM}} (z) = \frac{\mathrm{N_p^{Corr}}}{\mathrm{N_g^{Corr}}} =  \frac{\sum_{\mathrm{K=1}}^{\mathrm{N_p}} \mathrm{C_1}^{-1} \mathrm{C_2}^{-1} \omega_{\mathrm{K}}}{\sum_{\mathrm{i=1}}^{\mathrm{N_g}} \mathrm{C_i}^{-1}}.
\end{equation}
Here, $\mathrm{N_g}$ is the number of galaxies in the parent sample, $\mathrm{N_p}$ is the number of close pairs, $\mathrm{C_i}$ is the spectroscopic completeness of each mock galaxy as a function of magnitude and redshift where i=1,2 corresponds to each galaxy in an observed pair K, and $\omega_{\mathrm{K}}$ is the area correction factor accounting for both survey boundaries and the minimum projected distance. That is
\begin{equation}
\omega_{\mathrm{K}} = \frac{(\mathrm{r_p^{max}})^2}{(\mathrm{r_p^{max}})^2 -(\mathrm{r_p^{min}})^2} \times \frac{\pi (\mathrm{r_p^{max}})^2}{\mathrm{A_{_{JWST}}}}
\end{equation}
where $\mathrm{A_{_{JWST}}}$ is the survey area enclosing the circle of radius $\mathrm{r_p^{max}}$ centered on the more massive source in pair K. $\mathrm{N_g^{Corr}}$ and $\mathrm{N_p^{Corr}}$ are then the completeness corrected number of parent galaxies and major merger galaxy pairs, respectively. We compute pair fraction uncertainties via a jackknife approach to account for sample variance. As with the simulated clustering measurements, ten jackknife samples are obtained by splitting the footprints of each JADES fields into 5 roughly equal-area regions split by constant right ascension.

We show the simulated JADES/NIRSpec major merger pair fraction measurements at z$\sim$4-8 in Figure \ref{fig:MajorMergerPairFraction}. We also plot for comparison the MUSE measurements from \citetalias{Ventou2017} at z$\sim$0-6. We find that future \JWST{} surveys will substantially improve current galaxy pair fraction measurements at z$\sim$4-6 and establish the first measurements at z$\sim$7-8. Specifically, we find that errors on measured major merger pair fractions will be $\lesssim$0.1 dex from z$\sim$4-6 and $\sim$0.1-0.2 dex from z$\sim$7-8, made possible by the much wider coverage enabled by \JWST{}/NIRSpec ($>$100 arcmin$^2$) relative to MUSE ($\sim$10 arcmin$^2$). Given that we have not considered contributions from the JADES deep field nor pair fraction measurements made via photometry alone (c.f., \citealt{Duncan2019}), it is plausible that z$\sim$10 measurements will be possible within the first few cycles. We conclude that \JWST{} will clarify the importance of major mergers at z$\gtrsim$4 and into the epoch of reionization. 

\section{Conclusions} \label{sec:conclusions}
We have simulated \JWST{} galaxy clustering measurements at z$\sim$4$-$10 by adopting footprints and typical depths of the planned Cycle 1 GTO program, JADES \citep{Williams2018}, and utilizing an empirical model, the \textsc{UniverseMachine} \citep{Behroozi2019}, to assign galaxy properties to halos from a dark matter simulation. We have also assessed the ability of future \JWST{} surveys to quantify galaxy satellite fractions, identify individual satellites, and measure galaxy major merger pair fractions at z$\gtrsim$4. Conclusions from this study include:

\begin{enumerate}

\item Planned Cycle 1 \JWST{} surveys will measure galaxy angular clustering with $\gtrsim$5$\sigma$ significance at z$\sim$4$-$10. Dedicated spectroscopic follow-up over $\sim$150 arcmin$^2$ will enable $\sim$10$-$15$\sigma$ spatial clustering measurements at z$\sim$4$-$6 and $\sim$8$\sigma$ measurements at z$\sim$7.

\item Halo mass uncertainties resulting from Cycle 1 angular clustering measurements will be $\sim$0.2 dex for faint (-18 $\gtrsim$ \Muv{} $\gtrsim$ -20) galaxies at z$\sim$4$-$10 as well as $\sim$0.3 dex for bright (\Muv{} $\sim$ -20) galaxies at z$\sim$4$-$7. Dedicated spectroscopic follow-up over $\sim$150 arcmin$^2$ would yield $\sim$0.10$-$0.15 dex halo mass uncertainties for faint galaxies as well as $\sim$0.2 dex uncertainties for bright galaxies at z$\sim$4$-$6. Future \JWST{} observations will therefore provide the first constraints on the stellar-halo mass relation in the epoch of reionization and substantially clarify how this relation evolves with redshift at z$>$4.

\item Cycle 1 observations will allow precise inferences on galaxy satellite fractions at z$\sim$4$-$10 by enabling high-precision ($\sim$5$-$10$\sigma$) small-scale galaxy clustering measurements at these redshifts. \JWST{} observations will therefore soon test sub-halo occupation models in the early Universe.

\item It will be possible to identify $\sim$200 individual satellites at z$\sim$4$-$5 from Cycle 1 surveys. Furthermore, comprehensive follow-up of \HST{} legacy surveys with NIRSpec would enable the identification of $\sim$500 satellites at z$\sim$6$-$8. \JWST{} will therefore be able to test the environmental dependence of star formation in the early Universe.

\item Future \JWST{} surveys can substantially improve current galaxy major merger pair fraction measurements at z$\sim$4$-$6 and establish the first measurements at z$\gtrsim$7. Specifically, we find that dedicated NIRSpec follow-up over $\sim$150 arcmin$^2$ would yield $\lesssim$0.1 dex errors at z$\sim$4$-$6 and $\sim$0.1$-$0.2 dex at z$\sim$7$-$8. Such measurements can be used to quantify galaxy major merger rates and determine the relative role of mergers to the build-up of stellar mass into the epoch of reionization.
\end{enumerate}

\section*{Acknowledgements}

We thank Gurtina Besla and Rachel Somerville for stimulating discussions. R.E., D.P.S., and M.R. acknowledge funding from JWST/NIRCam contract to the University of Arizona, NAS5-02015. C.C.W. acknowledges support from the National Science Foundation Astronomy and Astrophysics Fellowship grant AST-1701546. B.E.R. acknowledges a Maureen and John Hendricks
Visiting Professorship at the Institute for Advanced Study,
NASA contract NNG16PJ25C, and NSF award 1828315. G.Y. acknowledges partial financial support under research grants AYA2015-63819-P (MINECO/FEDER) and PGC2018-094975-C21 (MICINN/FEDER).

The authors wish  to thank V. Springel for allowing us to use the L-Gadget2 code to run the different Multidark simulation boxes, including the VSMDPL used in this work. The VSMDPL simulation has been performed at LRZ Munich within the project pr87yi. The CosmoSim database (\url{www.cosmosim.org}) provides access to the simulation and the Rockstar data is a service by the Leibniz Institute for Astrophysics Potsdam (AIP). 

This research made use of Astropy, a community-developed core Python package for Astronomy \citep{astropy:2013, astropy:2018}; Matplotlib \citep{Hunter2007_matplotlib}; NumPy \citep{van2011numpy}; and SciPy \citep{jones_scipy_2001}.
 



\bibliographystyle{mnras}
\bibliography{paper_ref} 



\appendix

\section{Clustering Results When Using Abundance Matching to Assign UV Luminosities to Halos} \label{appendix:abundance_matching}

Here, we investigate how the clustering results shown in Figure \ref{fig:AutoCorrelations} would change if we instead used abundance matching to assign UV luminosities to halos within the UM-VSMDPL mock catalog. For the sake of consistency, we abundance match to the same empirical high-redshift UV luminosity functions as those adopted by the UM model, namely the \citet{Finkelstein2015_LF} luminosity functions at z$\sim$4$-$8 and the \citet{Bouwens2016_z910_LF} luminosity functions at z$\sim$9$-$10. We use the best-fit analytic Schechter form at each redshift to generate the galaxy counts as a function of \Muv{} and follow the abundance matching with scatter algorithm described in \citet{Allen2018}. Here, we order halos in the UM-VSMDPL mock catalog by their peak mass and adopt a fixed scatter in \Muv{} of $\sigma$ = 0.5 mags for simplicity. This value was chosen to be consistent with the scatter inferred by the UM model at high redshifts. All other methods for simulating clustering measurements remain as described in \S\ref{sec:methods} with the exception that the best-fit mock survey volumes (see \S\ref{sec:methods_survey_realizations}) are re-chosen using the UV luminosities assigned via abundance matching. While these best-fit volumes were originally selected using the same luminosity functions, the UV magnitudes assigned to each mock galaxy are different with abundance matching so we re-select the volumes for consistency.

\begin{figure*}
\begin{multicols}{2}
\includegraphics{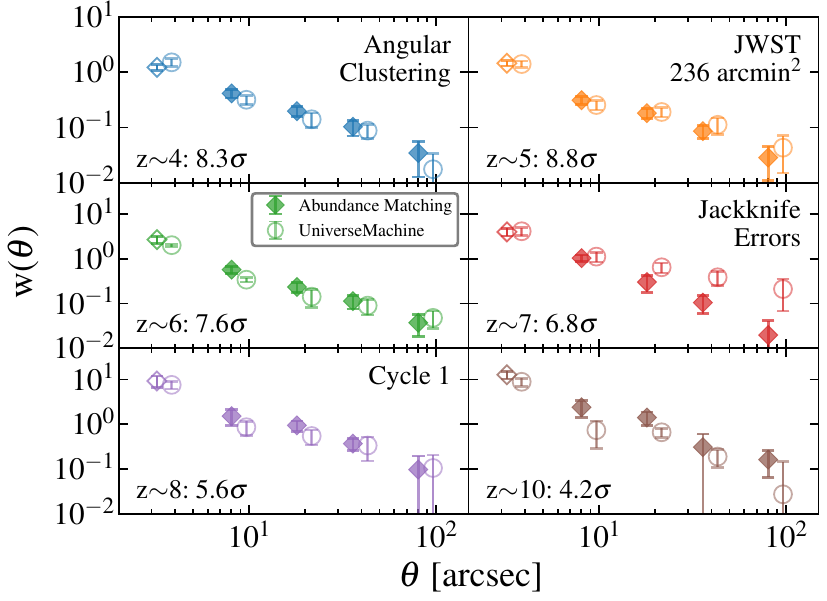}\par
\includegraphics{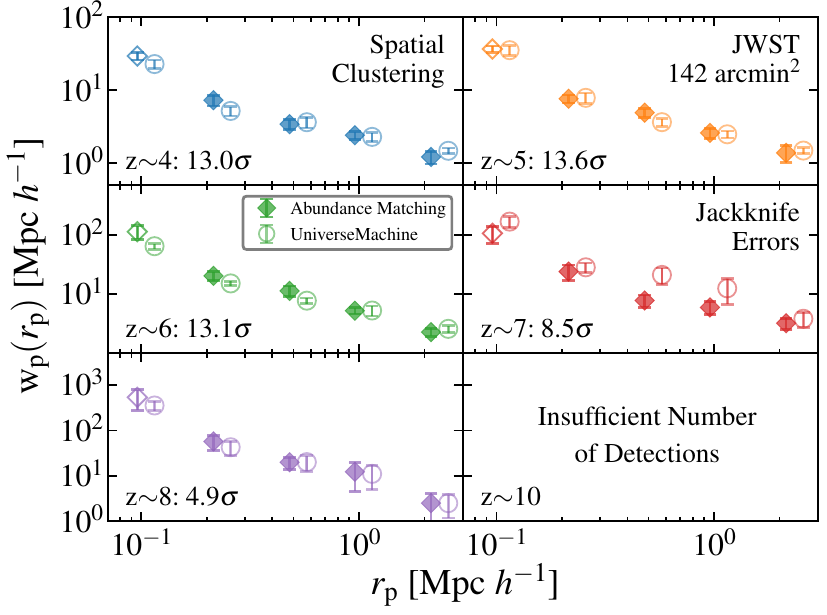}\par
\end{multicols}
\caption{Same as Figure \ref{fig:AutoCorrelations} except that we show typical simulated \JWST{} clustering measurements using \Muv{} values assigned via abundance matching with scatter with diamond markers. Original data points using \Muv{} values from the \textsc{UniverseMachine} model are shown with empty markers and are slightly offset to the right for clarity. In the lower left of each panel, we display typical measurement significances using \Muv{} values from abundance matching where it can be seen (by comparing to Figure \ref{fig:AutoCorrelations}) that these significances remain approximately the same between the two methods. Simulated clustering measurements are often lower when adopting the \textsc{UniverseMachine} model as satellites are assigned systematically fainter \Muv{} values relative to abundance matching.}
\label{fig:AutoCorrelations_compareAbundanceMatch}
\end{figure*}

As shown in Figure \ref{fig:AutoCorrelations_compareAbundanceMatch}, the typical clustering measurements and significances (again determined using simulated jackknife errors and ignoring the 1-halo term) obtained using abundance matching are consistent with those obtained using the UM model. The only particularly notable changes are at z$\sim$8 and z$\sim$10 where the measurement significances are $\sim$5$\sigma$. At z$\sim$8, the abundance matching approach yields slightly more optimistic measurement significances, moving from 5.0$\sigma$ to 5.6$\sigma$ for angular clustering and from 4.5$\sigma$ to 4.9$\sigma$ for spatial clustering. At z$\sim$10, the UM model approach yields a slightly higher typical measurement significance (4.9$\sigma$) than abundance matching (4.2$\sigma$). This is because the z$\sim$10 luminosity function inferred by the UM model is slightly higher than the best fit reported by \citet{Bouwens2016_z910_LF}, thereby lowering Poisson noise in pair counts.

We choose to emphasize results using the UM model for two reasons. First, our abundance matching approach does not account for possible evolution in the scatter of the \Muv{}$-$\Mhalo{} relation with redshift or halo mass. Secondly, abundance matching does not account for the expectation that satellites have systematically lower star formation rates than centrals at fixed halo mass due to environmental quenching \citep[e.g.,][]{vandenBosch2008,Kawinwanichakij2016}. This means that satellites are less likely to be detected when using the UM model and explains why clustering strengths obtained via the UM model tend to be lower. The exception to this trend is at z$\sim$5. At this redshift, the UV luminosity function inferred by the UM model is systematically lower than that reported by \citet{Finkelstein2015_LF} (see Figure \ref{fig:LF_comparison}) meaning that the UM model is assigning fixed \Muv{} values to higher mass halos which are more strongly clustered. 

\section{The Impact of z$\geq$4 UV Luminosity Function Uncertainties} \label{appendix:UVLF_Uncertainty}

We also test how our inferred clustering measurement significances depend on the z$\geq$4 luminosity functions adopted by the UM model. Rather than re-run the UM model with different input luminosity functions, we use the abundance matching with scatter approach described in Appendix \ref{appendix:abundance_matching}. This approach is justified by the similar results obtained between the two methods (Appendix \ref{appendix:abundance_matching}).

At z$\sim$4$-$8, the \citet{Bouwens2015_LF} luminosity functions are also commonly adopted in the literature. Galaxy number densities implied by this work are typically higher than those reported by \citet{Finkelstein2015_LF} which was adopted by the UM model. This suggests that using the \citet{Bouwens2015_LF} luminosity functions will result in lower clustering strengths because galaxies at fixed UV magnitude are being assigned to lower mass halos. However, we also expect cosmic variance to decrease for the same reason. Finally, Poisson noise should decrease due to higher overall pair counts. In the end, we find that z$\sim$4$-$8 clustering measurement significances either remain at very similar values or are boosted by $\sim$1$\sigma$.

We re-simulate the z$\sim$10 clustering strength measurements using the \citet{Bouwens2019_z910LF} and \citet{Oesch2018_z10LF} luminosity functions at z$\sim$9 and z$\sim$10, respectively. These luminosity functions are $\sim$0.3 dex lower than those reported by \citet{Bouwens2016_z910_LF} so we expect higher clustering strength measurements with larger cosmic and Poisson variance. We find that the typical z$\sim$10 clustering measurement significance drops to 3.0$\sigma$ as a result. While these luminosity functions give more pessimistic results, we still find that Cycle 1 surveys are likely to enable $\sim$5$\sigma$ clustering measurements at z$\sim$10 (\S\ref{sec:results_clusteringStrengths}).

\begin{table*}
\centering
\begin{tabular}{P{1cm}P{9.5cm}P{5.5cm}}
\hline
Redshift & Color Cuts & Non-Detection Criteria \\ 
\hline
\multirow{2}{*}{z$\sim$10} & F115W$-$F150W$>$1.6  $\wedge$  -2.15$<$F150W$-$F200W$<$0.55  $\wedge$ &  SN(F090W)$<$2  $\wedge$  SN(F070W)$<$2  $\wedge$  \\
                          & F115W$-$F150W$>$0.1(F150W$-$F200W)$+$1.6 & SN$_{\mathrm{IVW}}$(435,606,775)$<$2.5  \\
\hline
\multirow{2}{*}{z$\sim$8} & F090W$-$F115W$>$2  $\wedge$  F150W$-$F200W$<$0.4  $\wedge$ &  SN(F070W)$<$2  $\wedge$  SN$_{\mathrm{IVW}}$(435,606,775)$<$2.5 \\
                          & F090W$-$F115W$>$2.6(F090W$-$F115W)$+$2  $\wedge$ F410M$-$F444W$>$0 & $\wedge$  Not in z$\sim$10 selection \\
\hline
\multirow{2}{*}{z$\sim$7} & F090W$-$F115W$>$1.3  $\wedge$  F150W$-$F200W$<$0.4  $\wedge$ &  SN(F070W)$<$2  $\wedge$  SN$_{\mathrm{IVW}}$(435,606,775)$<$2.5 \\
                          & F090W$-$F115W$>$2.6(F090W$-$F115W)$+$1.3 & $\wedge$  Not in z$\sim$8-10 selection \\
\hline
\multirow{3}{*}{z$\sim$6} & F070W$-$F090W$>$1  $\wedge$  F115W$-$F150W$<$0.35  $\wedge$ &  SN(F435W)$<$2  $\wedge$  SN(F606W)$<$2  $\wedge$  \\
                          & F070W$-$F090W$>$2.4(F070W$-$F090W)$+$1  $\wedge$  F410M$-$F444W$>$-0.1  $\wedge$ & Not in z$\sim$7-10 selection \\
                          & F410M$-$F444W$>${}$-$1.16(F200W$-$F277W) $+$ 0.28 & \\
\hline
\multirow{2}{*}{z$\sim$5} & F090W$-$F115W$<$0.24  $\wedge$  F200W$-$F277W$>$0.28  $\wedge$ &  SN(F435W)$<$2  $\wedge$  \\
                          & F410M$-$F444W$>${}$-$1.06(F200W$-$F277W)$+$0.34  $\wedge$  F410M$-$F444W$>$-0.1 & Not in z$\sim$6-10 selection \\
\hline
\multirow{2}{*}{z$\sim$4} & F115W$-$F150W$<$0.25  $\wedge$  &  SN(F435W)$<$2  $\wedge$  \\
                          & F335M$-$F356W$<$1.92(F150W$-$F200W)$-$0.52 & Not in z$\sim$5-10 selection \\
\hline
\end{tabular}
\caption {Photometric redshift color cuts and non-detection criteria designed to be used with medium depth \JWST{} photometry along with deep optical \HST{} photometry. The SN$_{\mathrm{IVW}}$(435,606,775) values are the inverse-variance weighted signal-to-noise values of the F435W, F606W, and F775W optical \HST{} filters. All galaxies are also required to have a $\chi^2>25$ using all filters redder than and including F200W.} \label{tab:photzCuts}
\end{table*}

\section{Modeling Photometric Selection of High-Redshift Galaxies} 

\subsection{Determining Optimal Color Cuts} \label{appendix:photz_selec}

To ensure that our simulated high-redshift galaxy clustering measurements are realistic, we account for selection completeness. We assume that high-redshift galaxies will be photometrically selected with \JWST{} using color cuts as commonly done with \HST{} imaging \citep[e.g.,][]{Stark2011,Gonzalez2012,Oesch2014_z910LF,Bouwens2015_LF,Bouwens2019_z910LF}. To determine the most optimal color cuts for z$\sim$4-10 galaxy selection, we utilize the JAdes extraGalactic Ultradeep Artificial Realizations (JAGUAR; \citealt{Williams2018}) package which provides \HST{} and \JWST{} broadband photometry for mock galaxies from z=0.2 to z=15. The properties of the high-redshift JAGUAR mock galaxies are empirically constrained by stellar mass functions, stellar mass to UV luminosity relations, and UV luminosity to UV slope relations. See \citet{Williams2018} for further details. 

We first add noise to JAGUAR mock galaxy photometry 50 times for each source using the medium \JWST{} depths listed in Table 5 of \citet{Williams2018} and the deep GOODS-S \HST{} depths listed in Table 1 of \citet{Bouwens2015_LF}. Specifically, these (5$\sigma$) depths are m = 28.8, 29.4, 29.5, 29.7, 29.8, 29.4, 28.8, 29.4, 28.9, and 29.1 for the F070W, F090W, F115W, F150W, F200W, F277W, F335M, F356W, F410M, and F444W \JWST{} bands, respectively, and m = 28.7, 28.0, and 27.5 for the F435W, F606W, and F775W \HST{} bands, respectively. We then iterate through potential photometric redshift selection criteria in various redshift intervals at z$\sim$4-10 using a combination of Lyman-break, rest-UV continuum, rest-optical emission line, and Balmer-break color cuts, seeking to maximize selection completeness and minimize the fraction of low-redshift contaminants. 

We find that the color cuts listed in Table \ref{tab:photzCuts} are optimal. These color cuts separate galaxies into approximate photometric redshift bins of [3.7,4.4], [4.4,5.5], [5.5,6.7], [6.7,7.7], [7.7,9.0], and [9.0,11.0] which we refer to as the z$\sim$4, 5, 6, 7, 8, and 10 intervals, respectively. Within the context of JAGUAR, they successfully select $>$50\% of galaxies at apparent rest-UV magnitudes of \muv{} = 28.0, 28.5, 28.8, 29.2, 29.2, and 29.5 at z$\sim$4, 5, 6, 7, 8, and 10, respectively, while simultaneously introducing only 2-15\% fractions of low-redshift contaminants. We use these color cuts to simulate both photometric and spectroscopic completeness as described in \S\ref{sec:methods_selection_effects}.

\subsection{Accounting for Contaminants in Angular Clustering Measurements} \label{appendix:photz_contaminants}

After applying color cuts to the JAGUAR catalog \citep{Williams2018}, we find that low-redshift (z$<$3) contaminants will comprise $\sim$2$-$15\% of the photometrically selected z$\sim$4$-$10 galaxy samples used for simulated angular clustering measurements. We assume that these low-redshift contaminants will not be clustered with one another because none of the low-redshift contaminant distributions are sharply peaked in redshift space within the context of JAGUAR. We therefore insert a sample of randomly positioned objects into the angular clustering measurements so that they populate a fixed fraction of the total sample equal to our estimated low-redshift contamination fractions \citep{Williams2011}. Specifically, these fractions are 2.4\%, 9.1\%, 2.1\%, 5.3\%, 1.8\%, and 16.3\% at \zphot{}$\sim$4, 5, 6, 7, 8, and 10, respectively. Because the vast majority of these contaminants are within one magnitude of our adopted limiting magnitudes (listed in Table \ref{tab:survey_params}), we do not introduce these random points in the brighter threshold samples. 

There will also be high-redshift contaminants which lie slightly outside the targeted redshift interval in photometrically selected samples. These sources may be clustered with each other and galaxies which lie inside the targeted redshift interval. We therefore select high-redshift contaminants according to modeled photometric selection completeness as a function of redshift and magnitude using the color cuts in Table \ref{tab:photzCuts}. 

Unclustered contaminants decrease measured angular clustering strengths. For the case of low-redshift contaminants where each galaxy is completely unclustered relative to all other galaxies in the sample, auto-correlation clustering strengths are reduced by a factor of $\left(1 - f_{\mathrm{low}}\right)^2$ where $f_{\mathrm{low}}$ is the low-redshift contamination fraction. However, for the case of high-redshift contaminants, some of these sources may be clustered. We therefore treat the effective fraction of high-redshift contaminants as $f_{\mathrm{high,eff}} = X_{\mathrm{eff}} \times f_{\mathrm{high}}$ where $f_{\mathrm{high}}$ is the inferred fraction of high-redshift contaminants and $X_{\mathrm{eff}}$ is some value between 0 and 1. Adopting $X_{\mathrm{eff}} = 0$ assumes that the high-redshift contaminants are as equally clustered as the intended galaxy sample. Conversely, adopting $X_{\mathrm{eff}} = 1$ assumes that every high-redshift contaminant is completely unclustered with all other galaxies in the sample. We find that fixing $X_{\mathrm{eff}} = 0.75$ leads to angular clustering strength measurements that best match the `true' clustering strengths (\S\ref{sec:methods_twoptCF}) which only include galaxies inside the targeted redshift interval. Therefore, we take the total contamination fraction of each galaxy sample as $f_{\mathrm{low}} + 0.75 \times f_{\mathrm{high}}$. It will, in principle, be possible to more accurately forward model these effects once \JWST{} begins to supply a large sample of spectroscopic redshifts at z$>$4.

\section{Accounting for Spectroscopic Follow-Up Efficiency} \label{appendix:spec_followup}

In our simulated clustering measurements, we adopt the survey design of the planned Cycle 1 JADES \citep{Williams2018} program which includes 26 NIRSpec micro-shutter assembly (MSA) pointings. Every MSA pointing will be comprised of R$\sim$100 prism exposures as well as exposures with the R$\sim$1000 G235M and G395M grisms. Each R$\sim$100 MSA pointing can hold $\sim$200 sources (NIRSpec team private communication). Using the JAGUAR mock catalog \citep{Williams2018}, we find that the typical JADES R$\sim$100 prism depths will provide $\gtrsim$50\% spectroscopic redshift completeness for z$\sim$4-6 galaxies brighter than the limiting magnitudes set by our modeled photometric selection (Table \ref{tab:survey_params}). Applying this photometric selection to the UM-VSMDPL catalog (\S\ref{sec:methods_survey_realizations}), we expect that $\sim$5300 galaxies will be photometrically selected to be at z$\sim$4$-$6 over the 142 arcmin$^2$ of JADES/NIRSpec coverage. Therefore, the 26 MSA pointings are sufficient to follow-up nearly every z$\sim$4$-$6 candidate with R$\sim$100 spectroscopy assuming that each MSA pointing is completely filled with high-redshift galaxies. 

Due to the smaller dispersion, each R$\sim$1000 pointing will likely only be able to hold $\sim$75 sources (NIRSpec team private communication) suggesting a maximum of $\sim$2000 sources over 26 pointings. The JADES R$\sim$1000 depths will provide $\gtrsim$50\% spectroscopic redshift completeness for z$\sim$7 and 8 galaxies brighter than \muv{} = 29.0 and 28.7, respectively, within the context of JAGUAR. Applying modeled photometric selection to the UM-VSMDPL mock catalog, we expect $\sim$800 such z$\sim$7$-$8 candidates to be photometrically selected over the 142 arcmin$^2$ of JADES/NIRSpec coverage, well below the maximum $\sim$2000. We therefore assume that every z$\sim$7 and 8 photometric candidate brighter than our assumed spectroscopic limiting magnitudes (see Table \ref{tab:survey_params}) will also be spectroscopically followed up with NIRSpec. 

While the above calculations show that the 26 JADES MSA pointings can, in principle, hold targets equal to the number of z$\gtrsim$4 photometric candidates, we have not accounted for target placement restrictions to avoid overlapping spectra. Based on our experience with ground-based spectroscopy, a maximum of $\sim$50\% of available targets can plausibly be placed on a multi-object mask. We therefore discuss how the spatial clustering results presented in \S\ref{sec:results_clusteringStrengths} would change if we assume a 50\% spectroscopic follow-up efficiency for the JADES Cycle 1 program. Assuming that targets of different magnitudes are assigned the same follow-up priority, cosmic variance will not be altered. However, Poisson error will approximately double as a result because $\sigma_{\mathrm{Poisson}} \propto 1/\sqrt{D1D2} \propto 1/\Sigma$ \citep{Landy1993} where $\Sigma$ is galaxy surface density. This suggests that spatial clustering will be measured with similar (z$\sim$4$-$5) or weaker (z$\gtrsim$6) precision relative to angular clustering with the Cycle 1 JADES program alone. Additional dedicated spectroscopic follow-up will be necessary to achieve the $\sim$15$\sigma$ precisions quoted in Figure \ref{fig:AutoCorrelations}. Nonetheless, spatial clustering measurements are more desirable (if measured with similar significance) as they remove the systematic uncertainty associated with redshift contamination. 

\section{Satellite Evolution in the UniverseMachine Model} \label{appendix:satellite_evolution}

The \textsc{UniverseMachine} (hereafter UM; \citealt{Behroozi2019}) model accounts for both satellite and central halo evolution, allowing us to investigate not only central galaxy clustering, but satellite-satellite and satellite-central clustering as well. During the halo identification process, halos are identified as true satellites if they reside within the virial radius of a more massive halo. It is well known that multiple systematic effects lead to the premature disruption of low-mass satellites within dark matter simulations \citep[e.g.,][]{vandenBosch2018a,vandenBosch2018b}. The UM model compensates for these effects by allowing disrupted satellites to continue orbiting their host as orphan galaxies \citep{Wang2006} until their expected mass loss (computed via \citealt{Jiang16}) reaches a preset threshold. This threshold is tuned to match z=0-1 clustering constraints for star-forming and quiescent galaxies and it is assumed not to evolve with redshift. The net effect is to extend satellite lifetimes by $\sim$25\% at all redshifts; however, the total satellite fraction is always $\le$15\% at z$>$4. 

We note that our simulated galaxy major merger pair fraction measurements (\S\ref{sec:app_merger}) are sensitive to the orphan populations introduced by the UM model. Ignoring all orphans reduces the measured pair fractions by 0.3 dex at z$\sim$4 and 0.2 dex at z$\sim$5, which would make our simulated measurements less consistent with observations (Figure \ref{fig:MajorMergerPairFraction}). This suggests that the satellite correction methodology adopted by the UM model is reasonable at high redshifts and we therefore adopt this model as a plausible representation of satellite behavior at z$>$4 even as we acknowledge that clustering constraints from \JWST{} will inform better models in the future.

\section{Jackknife Errors on the Two-Point Correlation Function with JWST} \label{appendix:jackknife}

Jackknife and bootstrap methods are commonly used to determine the error on observationally measured two-point correlation functions. However, it is known that these errors should be interpreted with caution as they do not necessarily follow a Gaussian error distribution nor are they necessarily accurate \citep{Norberg2009}. On one hand, this is because clustering depends not only on halo mass but also environment, halo formation time, concentration, and subhalo occupation \citep[e.g.][]{Sheth2004,Gao2005,Wechsler2006,Pujol2014}. This phenomenon is broadly referred to as `assembly bias' and can be particularly strong in the narrow pencil-beam fields \JWST{} is designed to survey. Thus, if \JWST{} survey volumes are not sufficiently large to capture the full extent of clustering variations for a given galaxy sample, clustering strength errors will be underestimated. Furthermore, the number of subsamples into which the observational data are split must be sufficiently large to accurately derive the true errors. 

We test whether these concerns apply to future \JWST{} surveys by asking how often the `true' two-point correlation functions (\S\ref{sec:methods_twoptCF}) are enclosed by jackknife errors for the 100 simulated clustering measurement realizations for each galaxy sample (\S\ref{sec:methods_survey_realizations}). Jackknife errors are calculated by dividing each of the two GOODS fields into 5 roughly equal area segments split by constant right ascension for a total of 10 jackknife samples. We find that these jackknife errors have typical 1$\sigma$ and 2$\sigma$ confidence intervals of $\sim$70\% (i.e., jackknife errors enclose the true correlation functions in 70 of the 100 realizations) and 94\%, respectively, with JADES angular clustering measurements and $\sim$62\% and 88\%, respectively, with JADES spatial clustering measurements. Because these confidence intervals are not highly different from those expected of a Gaussian distribution and because we wish to predict observational clustering measurements, we adopt jackknife errors throughout this work.

\begin{figure}
\includegraphics{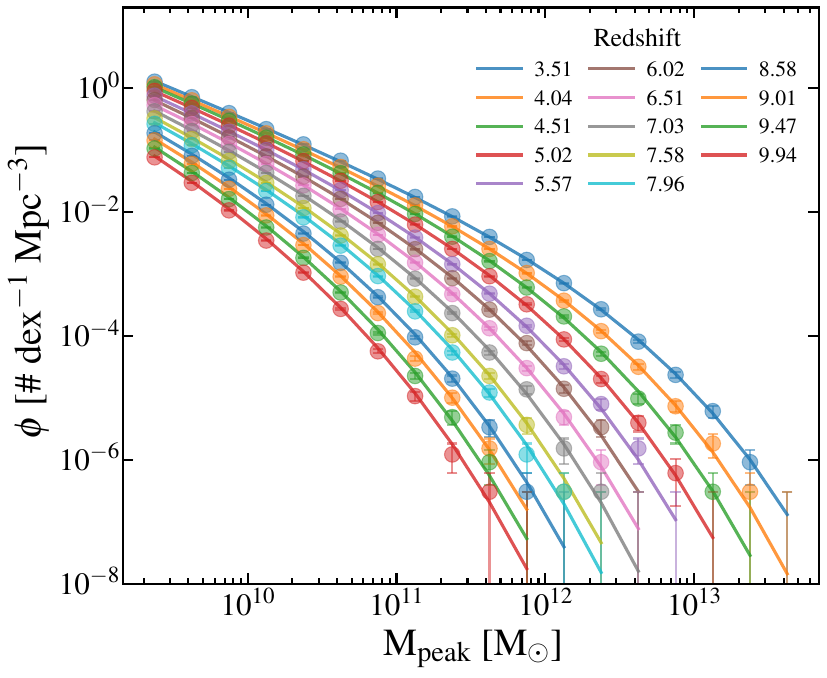}\par
\caption{Evolution in the peak halo mass (\Mpeak{}) function at z=3.5-10. Markers show data points from the UM-VSMDPL mock catalog (\S\ref{sec:methods_survey_realizations}) with Poisson errors. The lines show the fit from Equation \ref{eq:HMF_evol} at each redshift.}
\label{fig:HMF_Evolution}
\end{figure}

\section{Evolution of Peak Halo Mass Function at z$\sim$4-10} \label{appendix:HMF}

Here, we derive a smooth analytic fit to the redshift evolution of the peak halo mass function from z$\sim$4-10. We calculate the intrinsic peak halo mass function at snapshot redshifts spanning z=3.5-10 from the VSMDPL dark matter simulation (\S\ref{sec:methods_survey_realizations}) using only halos with masses of $\logten{} \left(\Mpeak{}/\Msol{}\right) > 9.5$. After experimentation with various functional forms, we find that the following form best matches the intrinsic evolution over these redshifts:
\begin{equation}
\label{eq:HMF_evol}
\begin{split}
\logten \phi \left(\Mpeak{},z\right) = & \logten \phi^{\ast} + \alpha \left(x - \Mst{}\right) \\
& - \logten\left(e\right) 10^{\ \beta \left(x - \Mst{}\right)}.
\end{split}
\end{equation}
Here $x \equiv \logten{} \left(\Mpeak{}/\Msol{}\right)$. We find that it is not necessary to leave $\alpha$ (which represents the faint-end slope) as a free parameter and fix it at $\alpha$ = $-$0.95. The evolutionary forms of the other parameters are found to be:
\begin{equation}
\beta \left(z\right) = -0.0185\ z + 0.488
\end{equation}
\begin{equation}
\begin{split}
\Mst{} \left(z\right) = & 10.944\ -\ 0.652\left(z-4.558\right) \\
& +\ 0.204\ \logten \left(1 + 10^{\ 0.820 \left(z-4.558\right)}\right)
\end{split}
\end{equation}
\begin{equation}
\begin{split}
\logten{} \phi^{\ast} \left(z\right) = & -1.361\ +\ 0.533\left(z-4.690\right) \\
& -\ 0.0714\ \logten \left(1 + 10^{\ 1.722 \left(z-4.690\right)}\right)
\end{split}
\end{equation}

This analytic evolutionary form agrees with the VSMDPL data at z=3.5-10 to within either 2$\sigma$ (Poisson) or 0.10 dex at each data point (binned by 0.25 dex in \Mpeak{}). These data points and the fit from Equation \ref{eq:HMF_evol} are shown in Figure \ref{fig:HMF_Evolution}. 

We note that these halo mass functions include the contribution from satellites identified by our adopted halo finder (\textsc{Rockstar}; \citealt{Behroozi2013_ROCKSTAR}) as well as orphan satellites that have been introduced to account for the premature disruption of low-mass satellites within dark matter simulations (Appendix \ref{appendix:satellite_evolution}). Excluding orphan satellites would have a very small impact on the intrinsic peak halo mass function at z=3.5-10, decreasing it by at most 0.04 dex. 


\bsp	
\label{lastpage}
\end{document}